\documentclass[12pt]{article}
\pdfoutput=1
\usepackage{jheppub}

\usepackage{amsmath,amsfonts,amssymb,amsthm,amssymb,bbm,bm}
\usepackage{pdfsync}
\usepackage{graphicx,import}
\usepackage[latin1]{inputenc}
\usepackage{hyperref}
\usepackage[all]{xy}
\usepackage{stmaryrd}
\usepackage{rotating}
\usepackage{color}  
\usepackage{slashed}
\usepackage{cancel}
\usepackage{array, makecell} %
\usepackage{comment}
\usepackage{tikz-cd}
\usepackage{hhline}
\usepackage{stackrel}

\newcommand{\bc}{\begin{center}}
\newcommand{\ec}{\end{center}}

\newcommand{\cO}{{\mathcal O}}

\newcommand{\be}{\begin{align}}
\newcommand{\ee}{\end{align}}
\newcommand{\bea}{\begin{eqnarray}}
\newcommand{\eea}{\end{eqnarray}}
\newcommand{\bs}{\begin{subequations}}
\newcommand{\es}{\end{subequations}}
\newcommand{\nn}{\nonumber}

\newcommand{\tl}{\tilde}
\def\p{\partial}

\newcommand{\ii}{\mathrm{i}}

\newcommand{\dd}{\mathrm{d}}

\newcommand{\pd}{\partial}

\def\eps{\epsilon}

\def\th{\theta}

\def\m{\mu}

\def\mn{{\mu\nu}}

\allowdisplaybreaks

\title{\boldmath \Large Multipole expansion of gravitational waves:\\
memory effects and Bondi aspects}

\author[a]{Luc Blanchet,}
\author[b]{Geoffrey Comp\`ere,}
\author[a]{Guillaume Faye,}
\author[c]{Roberto Oliveri,}
\author[d]{Ali Seraj}

\affiliation[a]{$\mathcal{G}\mathbb{R}\varepsilon{\mathbb{C}}\mathcal{O}$, Institut d'Astrophysique de Paris, UMR 7095, \\ \textit{CNRS \& Sorbonne Universit{\'e}, 98\textsuperscript{bis} boulevard Arago, 75014 Paris, France}}
\affiliation[b]{Universit\'{e} Libre de Bruxelles, %Centre for Gravitational Waves, 
\textit{International Solvay Institutes, CP 231, B-1050 Brussels, Belgium}}
\affiliation[c]{LUTH, Laboratoire Univers et Th\'eories, Observatoire de Paris\\CNRS, Universit\'e PSL, Universit\'e Paris Cit\'e, 5 place Jules Janssen, 92190 Meudon, France}
\affiliation[d]{School of Mathematical Sciences, Queen Mary University of London,\\Mile End Road, E1 4NS, United Kingdom}

\emailAdd{luc.blanchet@iap.fr}
\emailAdd{gcompere@ulb.be}
\emailAdd{faye@iap.fr}
\emailAdd{roberto.oliveri@obspm.fr}
\emailAdd{a.seraj@qmul.ac.uk}

\abstract{In our previous work, we proposed an algorithm to transform the metric of an isolated matter source in the multipolar post-Minkowskian approximation in harmonic (de Donder) gauge to the Newman-Unti gauge. We then applied this algorithm at linear order and for specific quadratic interactions known as quadratic tail terms. In the present work, we extend this analysis to quadratic interactions associated with the coupling of two mass quadrupole moments, including both instantaneous and hereditary terms. Our main result is the derivation of the metric in Newman-Unti and Bondi gauges with complete quadrupole-quadrupole interactions. We rederive the displacement memory effect and provide expressions for all Bondi aspects and dressed Bondi aspects relevant to the study of leading and subleading memory effects. 
Then we obtain the Newman-Penrose charges, the BMS charges as well as the second and third order celestial charges defined from the known second order and novel third order dressed Bondi aspects for mass monopole-quadrupole and quadrupole-quadrupole interactions.}

\arxivnumber{2303.07732}

\begin{document}

\maketitle
\flushbottom

\section{Introduction}
\label{sec:intro}

The displacement memory effect, \emph{i.e.}, the permanent change in the wave amplitude from before to after the passage of a gravitational wave strain, is a definite prediction of general relativity in the non-linear regime~\cite{Blanchet1990,Christodoulou:1991cr,Wiseman:1991ss,Th92,Blanchet:1992br}. The effect was initially computed using post-Newtonian (PN) theory in~\cite{Blanchet1990}, pointing out that it is dominantly of order 2.5PN in contrast to tails arising at 1.5PN order. This was published later, together with physical interpretation, in~\cite{Blanchet:1992br}. The effect can be interpreted as due to a finite accumulation of the electric (or mass type) part of the shear during the emission of gravitational radiation, which effectively induces a transition between the initial and final asymptotic Bondi-van der Burg-Metzner-Sachs (BMS)~\cite{1962RSPSA.269...21B,1962RSPSA.270..103S} frames related by a supertranslation~\cite{1966JMP.....7..863N,Strominger:2014pwa}. The amount of displacement memory caused by a compact binary has been computed analytically in the inspiral phase~\cite{Wiseman:1991ss,Blanchet:1992br,Blanchet:2008je} using post-Newtonian/post-Minkowskian (PN/PM) methods~\cite{Blanchet:1985sp,Blanchet:1986dk,Blanchet:1987wq} and, more recently, in the merger phase using numerical relativity with adapted asymptotic BMS frames~\cite{Mitman:2020pbt,Mitman:2021xkq}. Quite remarkably, it could be observed in the coming years with either ground-based gravitational-wave detectors~\cite{Favata:2009ii,Lasky:2016knh,McNeill:2017uvq} or pulsar timing arrays~\cite{Wang:2014zls,NANOGrav:2015xuc}.\footnote{Besides the non-linear memory effect of interest in this paper, the displacement memory is also caused at linear order in $G$ by a change of Bondi mass aspect (\emph{e.g.}, a change between initial and final compact object velocities)~\cite{Zeldovich:1974aa,Braginsky:1985vlg} and by matter radiation reaching null infinity~\cite{1977ApJ...216..610T,1978ApJ...223.1037E,Bieri:2010tq,Strominger:2014pwa}. These effects are usually called ordinary and null memory, respectively~\cite{Bieri:2013ada}}. 

Additional classes of memory effects, \emph{i.e.}, gravitational-wave observables that measure persistent changes between initial and final non-radiative states, have been constructed~\cite{Pasterski:2015tva,Nichols:2018qac,Flanagan:2018yzh,Himwich:2019qmj,Grant:2021hga,Seraj:2021qja,Seraj:2021rxd,Seraj:2022qyt,Godazgar:2022pbx} and their relationship to extended symmetry structures of asymptotically flat spacetimes~\cite{Barnich:2009se,Barnich:2010eb,Campiglia:2015yka,Compere:2017wrj,Godazgar:2018qpq,Godazgar:2018dvh,Compere:2019odm,Strominger:2021lvk} has been  partially established~\cite{Kapec:2016jld,Godazgar:2019dkh,Freidel:2021ytz}. Such memory effects are subleading in magnitude but could be potentially observed from space-borne gravitational-wave detectors~\cite{Nichols:2018qac}. 

Memory effects, especially the subleading ones, are most easily described in radiative gauges (see~\cite{Papa69,MadoreI,MadoreII} for a general definition), such as Bondi gauge~\cite{1962RSPSA.269...21B,1962RSPSA.270..103S} or closely related Newman-Unti (NU) gauge~\cite{newman1963class,Barnich:2011ty}. In this gauge-fixed formulation, the formal asymptotic Einstein solution can be obtained systematically to derive all flux-balance laws resulting from Einstein's constraint equations in terms of an asymptotic expansion~\cite{1962PhRv..128.2851S,10.2307/2415610,Tamburino:1966zz,1985FoPh...15..605W,Barnich:2010eb,Grant:2021hga,Freidel:2021dfs}. Such flux-balance laws precisely encode the memory effects and their associated conserved charges~\cite{Barnich:2011mi,Flanagan:2015pxa,Barnich:2019vzx,Compere:2019gft,Mitman:2020pbt,Grant:2021hga}. 

In our joint recent work~\cite{Blanchet:2020ngx}, building upon the formulation of general radiative gauges with polynomial expansions~\cite{Blanchet:1986dk}, we constructed the algorithm to transform the metric of an isolated matter source in the multipolar post-Minkowskian (MPM) approximation in harmonic/de Donder gauge to NU gauge and we applied this algorithm at linear order and for specific quadratic interactions known as quadratic tail terms~\cite{Blanchet:1987wq}. The main technical objective of the present work is to derive the NU metric of such isolated matter source for the quadratic interaction associated with the coupling of two mass quadrupole moments. It is an intermediate milestone for achieving the larger aim of deriving the NU metric that entirely encodes the $2.5$PN 
%motion 
 radiation from binary compact mergers. This metric only requires tails, the quadrupole-quadrupole interactions which will be investigated here, and spin-quadrupole interactions \cite{Blanchet:1996pi,Arun:2004ff,Kidder:2007gz}. 

The quadrupole-quadrupole interaction provides the leading contribution to displacement memory, so this will allow us to discuss this effect as well as subleading memories within the NU metric setup for this interaction. In addition, we shall derive (starting from the result in harmonic coordinates) the complete quadrupole-quadrupole NU metric including the memory and all instantaneous (\emph{i.e.}, ``non-hereditary'') terms. In particular, we shall explicitly verify the cancellation of all far-zone logarithms associated with harmonic coordinates, which confirms the soundness of the general algorithm proposed in~\cite{Blanchet:2020ngx}.

The content of the paper is as follows. In section~\ref{sec:quadratic}, we review the presentation of hereditary terms in the MPM formalism following~\cite{Blanchet:1992br} and we discuss the Poincar\'e flux-balance laws using the updated results of~\cite{Nichols:2017rqr,Blanchet:2018yqa,Compere:2019gft}. We then apply the algorithm developed in~\cite{Blanchet:2020ngx} to memory and mass-loss hereditary terms arising at quadratic order (the remaining tail hereditary terms were treated previously in~\cite{Blanchet:2020ngx}), specializing to mass quadrupole-quadrupole interactions only when presenting explicit expressions. The result is valid to any order in the distance to the source. We find the corresponding Newman-Unti metric and Bondi data. We proceed with rederiving the Christodoulou formula for the displacement memory~\cite{Christodoulou:1991cr} using the energy flux-balance law. We finally compute the full quadratic NU metric corresponding to mass monopole-quadrupole and quadrupole-quadrupole interactions. 
In section~\ref{sec:IR}, we discuss various aspects of gravitational charges in the multipolar expansion. 
We first extend to the radiative case (but for the specific multipolar interactions considered) the classic result \cite{Newman:1965ik,10.2307/2415610} establishing that the Newman-Penrose charges \cite{Newman:1965ik,Newman:1968uj} can be expressed in the center-of-mass frame in terms of the product of the ADM mass and the initial mass quadrupole moment. 
Second, we extend the dictionary between NU gauge and Bondi gauge initiated in \cite{Blanchet:2020ngx} to quadratic order in $G$. We revise the flux-balance laws, the BMS and conserved celestial charges following recent developments \cite{Flanagan:2015pxa,Nichols:2018qac,Godazgar:2018qpq,Godazgar:2018dvh,Compere:2018ylh,Compere:2019gft,Godazgar:2019dkh,Barnich:2019vzx,Grant:2021hga} and provide a corrected definition of the dressed $n=3$ Bondi aspect alternative to  \cite{Grant:2021hga}, which was subsequently revised in \cite{PhysRevD.107.109902}. We proceed by explicitly computing all lower order $n=0,1,2,3$ Bondi charges in the presence of mass monopole-quadrupole and quadrupole-quadrupole interactions. We conclude in section~\ref{Sec:concl}.
Appendix~\ref{app:A} is devoted to the technical treatment of hereditary integrals. 
Appendix~\ref{app:B} gives more information on dressed Bondi aspects for quadratic interactions.

\paragraph{Notation and conventions}

We adopt units in which the speed of light $c$ is set to 1. The Newton gravitational constant $G$ is kept explicit to bookmark post-Minkowskian (PM) orders. Lower case Latin indices from $a$ to $h$ will refer to indices on the two-dimensional sphere, while lower case Latin indices from  $i$ to $z$ will refer to three-dimensional Cartesian indices. The Minkowski metric is $\eta_{\mu\nu}=\text{diag}(-1,+1,+1,+1)$.

We denote Cartesian coordinates as $x^{\mu} = (t, \mathbf{x})$ and spherical ones as $(t, r, \theta^a)$. More precisely, the radial coordinate $r$ is defined as $r=\vert\mathbf{x}\vert$, while $\theta^a=(\theta, \varphi)$ with  $a,b,\dots =\{1,2\}$. The unit directional vector is denoted as $n^i = n^i(\theta^a) = x^i/r$. Euclidean spatial indices $i, j, \dots = \{1,2,3\}$ are raised and lowered with the Kronecker metric $\delta_{ij}$.

Furthermore, $L=i_1i_2 \cdots i_\ell$ represents a multi-index made of $\ell$ spatial indices. We use short-hands for: the multi-derivative operator $\partial_L = \partial_{i_1}\cdots\partial_{i_\ell}$ with $\partial_i=\partial/\partial x^i$, the product of vectors $n_L=n_{i_1}\cdots n_{i_\ell}$, as well as $x_L=x_{i_1}\cdots x_{i_\ell}=r^\ell n_{L}$. The multipole moments $M_L$ and $S_L$ are symmetric-trace-free (STF) tensors over $L$. Time derivatives are indicated by superscripts $(q)$ or by dots. 

We define the Minkowskian outgoing vector $k^{\mu} \partial_{\mu} = \partial_{t} + n^i \partial_i$, or, in components, $k^{\mu} = (1, n^i)$ and $k_{\mu} = (-1, n^i)$. In retarded spherical coordinates $(u,r,\theta^a)$ with $u=t-r$, we have $k^\mu \partial_\mu = \partial_r\big\vert_{u}$. We commonly employ the natural basis on the unit 2-sphere embedded in $\mathbb R^3$, $e_a=\frac{\pd}{\pd \theta^a}$, with components $e^i_a = \partial n^i/\partial\th^a$. Given  the metric on the unit sphere $\gamma_{ab} = \mathrm{diag}(1,\sin^2\th)$, we have: $n^i e^i_a = 0$, $\partial_i \theta^a = r^{-1} \gamma^{ab} e_b^i$, $\gamma_{ab} = \delta_{ij} e^i_a e^j_b$ and $\gamma^{ab} e^i_a e^j_b  = \perp^{ij}$, where $\perp^{ij}=\delta^{ij}-n^i n^j$ is the projector onto the sphere tangent bundle. We also use the notation $e_{\langle a}^i e_{b\rangle}^j=e^i_{( a}e^j_{b)}-\frac{1}{2}\gamma_{ab}\!\perp^{ij}$ for the trace-free product of basis vectors, where round brackets denote symmetrization $e^i_{( a}e^j_{b)} = \frac{1}{2}(e^i_{a}e^j_{b}+e^i_{b}e^j_{a})$. On the other hand, square brackets denote anti-symmetrization, \emph{e.g.}, $e^i_{[a}e^j_{b]} = \frac{1}{2}(e^i_{a}e^j_{b}-e^i_{b}e^j_{a})$. 
 The transverse-trace-free (TT) projection operator is defined as $\perp^{ijkl}_\mathrm{TT} = \perp^{k(i}\perp^{j)l} - \frac{1}{2}\!\perp^{ij}\perp^{kl}$. Introducing the covariant derivative $D_a$ compatible with the sphere metric, $D_a \gamma_{bc}=0$, we can write $D_a e^i_b = D_b e^i_a = D_a D_b n^i = - \gamma_{ab} n^i$.  It is also convenient to denote $\Delta = D_c D^c$. An arbitrary rank 2 STF tensor over the sphere may be decomposed as 
\begin{equation}\label{decompT}
T_{ab}=-2 D_{\langle a} D_{b \rangle} T^+ +2 \eps_{c (a} D_{b)} D^c T^- \,,
\end{equation}
where $\eps_{ab}\equiv e_a^i e_b^j n_k \eps_{kij}$ is the unit sphere volume form and the modal decomposition of $T^\pm=\sum_{\ell \geqslant 2} T_L^\pm n_L$, with the $T_L^\pm$ being STF over the multi-index $L$, contains only $\ell \geqslant 2$ harmonics without loss of generality. Note the properties $D_{\langle a}e_{b \rangle}^i=0$, $D^a(e_{\langle a}^i e_{b \rangle}^j)=-2 n^{(i}e^{j)}_b$,  $\Delta T^\pm=-\ell(\ell+1)n_L T_L^\pm$, and $(\Delta+2) (e_{\langle a}^i e_{b\rangle}^j T^+_{ij})=0$. We define the unit-normalized integral over the sphere as $\oint_S  \, 1 = 1$.  

Given a general manifold, harmonic/de Donder coordinates are specified by using a tilde: $\tl x^{\mu} = (\tl t, \mathbf{\tl x})$ or $(\tl t, \tl r, \tl \theta^a)$. The metric components are $\tl g_{\mu\nu}(\tl x)$. Asymptotically flat spacetimes admit, as a background structure, the Minkowskian outgoing vector $\tl k^{\mu} = (1, \tl n^i)$, the basis on the sphere $\tl e^i_a = \partial \tl n^i/\partial\tl \th^a$, \textit{etc}. We define the retarded time $\tl u$ in harmonic coordinates as $\tl u=\tl t- \tl r$, so that $\tl k_\mu = - \tl \partial_{\mu} \tl u$.

NU coordinates are denoted as $x^{\mu} = (u, r, \theta^a)$ with $\theta^a=(\theta, \varphi)$. The metric components in those coordinates are $g_{\mu\nu}(x)$, with all other notations, such as the natural basis on the sphere $e^i_a$ and the metric $\gamma_{ab}$, as previously.

\section{Quadratic memory from the MPM formalism}
\label{sec:quadratic}

\subsection{Original formulation: modified harmonic coordinates}
\label{sec:modharmcoord}

Following the MPM formalism, the metric outside an isolated matter system is written as the formal post-Minkowskian (PM) expansion $h^\mn =\sqrt{|\tl g|}\,\tl{g}^\mn-\eta^\mn = G\,h_{1}^\mn + G^2\,h_{2}^\mn + \mathcal{O}(G^3)$, with each PM coefficient generated in the form of a multipole expansion starting from the linearized vacuum metric in harmonic coordinates~\cite{SB58,Pi64,Thorne:1980ru,1986RSPTA.320..379B} (setting $c=1$)
\begin{subequations}\label{eq:linearmetric}
	\begin{align}
		h_{1}^{00} &= -4 \sum_{\ell=0}^{+\infty}\frac{(-)^{\ell}}{\ell!}\tl\p_L\left(\frac{M_{L}(\tl u)}{\tl r} \right)\,, \\
		h_{1}^{0i} &= 4\sum_{\ell=1}^{+\infty} \frac{(-)^{\ell}}{\ell!}\biggl[\tl\p_{L-1}\biggl(\frac{\stackrel{(1)}{M}\!{}_{\!i L-1}(\tl u)}{\tl r} \biggr) + \frac{\ell}{\ell+1}\tl\p_{p L-1}\biggl(\frac{\eps_{ipq}S_{q L-1}(\tl u)}{\tl r} \biggr)\biggr]\,, \\
		h_{1}^{ij} &=  -4\sum_{\ell=2}^{+\infty} \frac{(-)^{\ell}}{\ell!}\biggl[\tl\p_{L-2}\bigg(\frac{\stackrel{(2)}{M}\!{}_{\!ij L-2}(\tl u)}{\tl r} \biggr) +  \frac{2\ell}{\ell+1}\tl\p_{p L-2}\biggl(\frac{\eps_{pq(i}\stackrel{(1)}{S}\!{}_{\!j)q L-2}(\tl u)}{\tl r} \biggr)\biggr]\,.
	\end{align}
\end{subequations} 
Here $M_L$ and $S_L$ are the symmetric-trace-free (STF) canonical mass and current multipole moments, which depend on the harmonic coordinate retarded time $\tl u = \tl t - \tl r$. For the following, it is convenient to write the dominant $1/\tl r$ piece of $h_1^\mn$ (when $\tl r\to+\infty$ with fixed $\tl u$) as\footnote{This tensorial quantity is presented in the form $h^{\mu\nu}= \left( \begin{array}{c} h^{00} \\ h^{0i} \\ h^{ij}\end{array}\right)$.}
\begin{equation}
\label{polar}
\hspace{-2.2cm}h_{1}^\mn = \frac{1}{\tl r}
\left(\begin{array}{c}-4\bigl( M + \tl n_i P^i\bigr) + z_{1}^{00}(\tl u, \tl {\bm{n}}) \\[0.3cm]-4 P^i + z_{1}^{0i}(\tl u, \tl {\bm{n}}) \\[0.3cm]z_{1}^{ij}(\tl u, \tl {\bm{n}}) 
\end{array}\right) + \mathcal{O}\left(\frac{1}{\tl r^2}\right)\,,
\end{equation}
where $M$ is the constant 
Arnowitt-Deser-Misner (ADM) energy, $S_i$ is the constant ADM angular momentum and $P^i \equiv M^{(1)}_i$ the constant ADM linear momentum of the system, $M_i$ being the mass dipole, identical for the conservative dynamics to the vector $G_i$ that defines the center-of-mass position (after division by $M$) and reduces to zero in the center-of-mass frame. We have introduced the quantities
\begin{subequations}\label{eq:zmunu}
	\begin{align}
	z_{1}^{00} &= -4 \sum_{\ell=2}^{+\infty}\frac{1}{\ell!}\,\tl n_L \!\stackrel{(\ell)}{M}\!{}_{\!L}(\tl u) \,, \\
	z_{1}^{0i} &= - 4 \sum_{\ell=2}^{+\infty} \frac{1}{\ell!}\left(\tl n_{L-1} \!\stackrel{(\ell)}{M}\!{}_{\!i L-1}(\tl u) + \frac{\ell}{\ell+1}\tl n_{p L-1} \eps_{ipq} \!\stackrel{(\ell)}{S}\!{}_{\!q L-1}(\tl u) \right)\,, \\
	z_{1}^{ij} &=  -4\sum_{\ell=2}^{+\infty} \frac{1}{\ell!} \left( \tl n_{L-2} \!\stackrel{(\ell)}{M}\!{}_{\!ij L-2}(\tl u) +  \frac{2\ell}{\ell+1} \tl n_{p L-2} \eps_{pq(i} \!\stackrel{(\ell)}{S}\!{}_{\!j)q L-2}(\tl u) \right)\,.
	\end{align}
\end{subequations} 
Note that $\tl k_\nu z_1^\mn = 0$ as the consequence of the harmonic gauge condition $\tl\p_\nu h_1^\mn = 0$, where we have introduced the Minkowski outgoing null vector $\tl k^\mu = (1, \tl n^i)$. At quadratic order, $h_{2}^\mn$ obeys the flat wave equation $\tl\Box h_{2}^\mn = N_{2}^\mn$ (together with $\tl\p_\nu h_{2}^\mn = 0$), where the source term $N_{2}^\mn$ is quadratic in $h_{1}$ as well as its first and second space-time derivatives. 

Hereditary terms with respect to the canonical moments $M_L$, $S_L$ are defined as retarded time integrals involving those moments; for instance tail terms are hereditary terms with typically some logarithmic kernel, and memory terms are given by time anti-derivatives of product of moments. It is known~\cite{Blanchet:1992br} that, at quadratic order, hereditary terms with respect to the canonical moments $M_L$, $S_L$ are generated only from the $1/\tl r^2$ piece in $N_{2}^\mn$, 
\begin{align}\label{eq:N2munu}
N_{2}^\mn &= \frac{1}{\tl r^2} \,Q_{2}^\mn(\tl u, \tl {\bm{n}}) + \mathcal{O}\left(\frac{1}{\tl r^3}\right)\,,
\end{align}
and that this piece takes the form
\begin{align}\label{eq:Q2munu}
Q_{2}^{\mu\nu}(\tl u, \tl {\bm{n}}) = 4 \bigl( M + \tl n_i P_i\bigr) \frac{\dd^2 z^\mn}{\dd \tl u^2} + \tl k^\mu \tl k^\nu \,\Pi(\tl u, \tl {\bm{n}})\,.
\end{align}
The first term will generate the tails while the second term is responsible for the displacement memory. The latter takes the form of the stress-energy tensor of gravitons propagating along the null direction $\tl k^\mu = (1, \tl n^i)$. The quantity $\Pi$ is directly given by a quadratic product of the multipole expansion~\eqref{eq:zmunu}:
\begin{align}\label{eq:defPi}
\Pi = \frac{1}{2} \frac{\dd z^\mn}{\dd \tl u}\frac{\dd z_\mn}{\dd \tl u} - \frac{1}{4} \frac{\dd z^\mu_\mu}{\dd \tl u}\frac{\dd z^\nu_\nu}{\dd \tl u}\,.
\end{align}
The GW energy flux emitted in the solid angle $\dd\tl \Omega$ around the direction $\tl{\bm{n}}$ is
\begin{align}\label{eq:energidensity}
\frac{\dd E^\text{GW}}{\dd\tl u\dd\tl\Omega} = \frac{G}{16\pi}\,\Pi(\tl u, \tl {\bm{n}}) + \mathcal{O}(G^2)\,.
\end{align}

The systematic study of hereditary terms (either tails or memory) at quadratic order was done in~\cite{Blanchet:1992br} with such formalism. In addition to the tail and memory terms, there are other hereditary terms in the harmonic gauge at this order. They are associated with GW losses of energy, linear and angular momenta, as well as center-of-mass position. The leading $1/\tl r$ components of $h_2^{\mu\nu}$ for multipole-multipole interactions in a radiative metric are given in eq.~(2.39) of~\cite{Blanchet:1992br}. Moreover, the hereditary terms in the subleading piece $1/\tl r^2$ are easily computed from~(2.29-30) of that paper. We can write
\begin{align}\label{eq:h2hered}
h_2^\mn{\Big|}_\text{hered} = h_2^\mn{\Big|}_\text{tail} + h_2^\mn{\Big|}_\text{mem}\,.
\end{align}
The tail term comes directly from the first term in eq.~\eqref{eq:Q2munu}; at the dominant order $1/\tl r$, it involves a logarithmic kernel,
\begin{align}\label{eq:h2tail}
h_2^\mn{\Big|}_\text{tail} = \frac{2(M+\tilde n_i P_i)}{\tilde r}\int_{-\infty}^{\tl u} \dd v \ln\left(\frac{\tl u - v}{2b_0}\right)\frac{\dd^2 z^\mn}{\dd \tl u^2}(v, \tl{\bm{n}}) + \mathcal{O}\left(\frac{1}{\tl r^2}\right)\,.
\end{align}
It will not be considered here since it has already been investigated in the previous paper~\cite{Blanchet:2020ngx}. The memory term actually contains both contributions mentioned above, from the non-linear memory strictly speaking and from the GW secular losses. For simplicity, we will just refer to these two types of terms as ``memory''. The memory is then given at quadratic order by the ``exact expression'', \emph{i.e.}, valid at any order in $1/\tl r$:
\begin{align}\label{eq:h2mem}
h_2^\mn{\Big|}_\text{mem} = \frac{1}{\tilde r}\int_{-\infty}^{\tl u} \dd v \,K^\mn(v, \tl{\bm{n}}) + \frac{1}{\tilde r^2}\left(\begin{array}{c} \tl n_i\biggl[\dfrac{1}{3}\!\stackrel{(-2)}{\Pi_i} + 4\!\stackrel{(-1)}{\Lambda_i} \biggr]\\[0.1cm]-2 \eps_{ipq} \,\tl n_p\!\stackrel{(-1)}{\Sigma_q}\\[0.1cm]0
\end{array}\right)\,.
\end{align}
The radiative coordinates used in this formula are defined in~\cite{Blanchet:1986dk} and in eqs.~(2.17)-(2.19) of~\cite{Blanchet:1992br}; they are not Newman-Unti coordinates but, since they are designed to remove all the $\ln \tl r$ components of the radiative metric to quadratic order, they belong, like the NU coordinates, to the large class of radiative coordinate systems~\cite{Papa69,MadoreI,MadoreII}. Although these coordinates are not harmonic either (we may call them modified harmonic coordinates), we will still use tildes to denote them, $\tilde x^\mu=(\tl u, \tl r, \tl{\bm{n}})$. 

The leading term in $1/\tl r$ is just a time anti-derivative (sometimes called ``semi-hereditary''). It is a combination of the memory and also the secular mass loss $\propto\Pi_0$ in the 00 component of $K^\mn$. To define it, we decompose the quantity $\Pi$ into STF multipolar pieces as
\begin{subequations}\label{eq:PiSTF}
	\begin{align}
	\Pi(\tl u, \tl{\bm{n}}) &= \sum_{\ell=0}^{+\infty} \tl n_L \,\Pi_L(\tl u)\,,\\
	\text{where}\quad\Pi_L(\tl u) &= \frac{(2\ell+1)!!}{\ell!} \!\int\frac{\dd\tl\Omega}{4\pi}\,\tl n_{\langle L\rangle}\,\Pi(\tl u, \tl{\bm{n}})\,.
	\end{align}
\end{subequations}
Then, the components of $K^\mn$ are explicitly given by\footnote{We corrected a sign typo in the last term of eq.~(2.40b) of~\cite{Blanchet:1992br}.} 
\begin{subequations}\label{eq:Kmunu}
	\begin{align}
	K^{00}(\tl u, \tl{\bm{n}}) &= \frac{1}{2}\Pi_0+\frac{1}{3}\tilde n_i \,\Pi_i , \\
	K^{0i}(\tl u, \tl{\bm{n}}) &= \frac{1}{3}\Pi_i+\frac{1}{2}\sum_{\ell \geqslant 0}\frac{1}{\ell+1} \tilde n_{iL}\,\Pi_L - \frac{1}{2}\sum_{\ell \geqslant 1}\frac{1}{\ell+1} \tilde n_{L-1}\,\Pi_{iL-1}, \\ 
	K^{ij}(\tl u, \tl{\bm{n}}) &= \sum_{\ell \geqslant 0}\frac{1}{\ell+2}\tilde n_{ijL}\,\Pi_L -2\sum_{\ell \geqslant 2}\frac{1}{(\ell+1)(\ell+2)}\tilde n_{L-2}\,\Pi_{ijL-2}\nn\\&-\sum_{\ell \geqslant 1}\frac{1}{(\ell+1)(\ell+2)}\Bigl[\delta_{ij}\tilde n_L \,\Pi_L + (\ell-2)\tilde n_{L-1(i}\,\Pi_{j)L-1}\Bigr]\,.\label{eq:Kij}
	\end{align}
\end{subequations}
Notice the useful properties, posing $K\equiv\eta_\mn K^\mn$:
\begin{equation}
\tl k_\nu K^\mn = 0\,,\qquad K = -\frac{1}{3} \tl n_i \,\Pi_i\,.     
\end{equation}
To order $1/\tl r$, the memory contribution to the transverse-traceless (TT) asymptotic waveform reads
\begin{align}\label{eq:TTWF}
H_{ij}^\text{TT}{\Big|}_\text{mem} &= - {\perp}{}^\text{TT}_{ijkl} \int_{-\infty}^{\tl u} \dd v \,K^{kl}(v, \tl{\bm{n}}) %\nn\\&= 
=\,{\perp}{}^\text{TT}_{ijkl} \sum_{\ell \geqslant 2}\frac{2\tl n_{L-2}}{(\ell+1)(\ell+2)} \int_{-\infty}^{\tl u} \dd v \,\Pi_{klL-2}(v) \,,
\end{align}
since all terms in $K^{kl}$ proportional to either $\tl n^k$, $\tl n^l$ or $\delta^{kl}$ are killed by the TT projection. The global minus sign arises from the relation $g^\text{TT}_{ij}=-h^\text{TT}_{ij}$, where $h_{ij}$ is the perturbation of the gothic metric (following the convention of~\cite{Blanchet:2020ngx}).

As for the subleading $1/\tl r^2$ piece in eq.~\eqref{eq:h2mem}, it is associated with GW secular losses; it contains both simple and double anti-derivatives indicated by the superscripts $(-n)$. We assume that the metric was stationary before some remote date in the past (for any $\tl u\leqslant -\mathcal{T}$). In this situation, all anti-derivatives are well defined.

From eq.~\eqref{eq:energidensity}, the angular average $\Pi_0 = \int\frac{\dd\tl\Omega}{4\pi}\,\Pi$ is proportional to the total energy flux carried by gravitational waves:
\begin{align}\label{fluxE}
\frac{\dd E^\text{GW}}{\dd\tl u}&= \frac{1}{4}G\,\Pi_0 + \mathcal{O}(G^2) \,.
\end{align}
Similarly, $\Pi_i = 3\int\frac{\dd\tl\Omega}{4\pi}\,\tl n_i\,\Pi$ is proportional to the total linear momentum flux. We also introduced the notation $\Sigma_i$ for the angular momentum flux and $\Lambda_i$ for the center-of-mass flux: 
\begin{align}\label{fluxes}
\frac{\dd P^\text{GW}_i}{\dd\tl u} = \frac{G\,\Pi_i}{12} +\mathcal{O}(G^2)\,,\quad \frac{\dd S^\text{GW}_i}{\dd\tl u} = G\,\Sigma_i+\mathcal{O}(G^2) \,,\quad \frac{\dd G^\text{GW}_i}{\dd\tl u} - P^\text{GW}_i =  G\,\Lambda_i +\mathcal{O}(G^2) \,.
\end{align}
We can write flux-balance equations equating the fluxes to losses in the source; see eqs. (2.29)-(2.30) of \cite{Blanchet:1992br}. The multipole decompositions of the fluxes are well known~\cite{Thorne:1980ru,1975ApJ...197..717E}, except for the one associated with the position of the center-of-mass, obtained only recently from the asymptotic properties of the radiation field~\cite{KQ16,KNQ18}, using the traditional PN as well as the Bondi approaches \cite{Nichols:2017rqr,Blanchet:2018yqa,Nichols:2018qac,Compere:2019gft}. The complete multipole expansions of the fluxes to quadratic order ($\propto G^2$) are given by~\footnote{ Fluxes are defined with the opposite sign as the corresponding ADM quantities. Accordingly, the fluxes of Poincar\'e charges $\mathcal E$, $\mathcal P_i$, $\mathcal J_i$ and $\mathcal G_i$ defined from canonical methods \cite{Compere:2019gft} are given in the notations used here as $\dot{\mathcal E}=-\frac{\dd E^{GW}}{\dd\tilde u}$, $\dot{\mathcal P}_i=-\frac{\dd P_i^{GW}}{\dd\tilde u}$, $\dot{\mathcal J}_i=-\frac{\dd S_i^{GW}}{\dd\tilde u}$ and $\dot{\mathcal G}_i- \mathcal P_i=-\frac{\dd G_i^{GW}}{\dd\tilde u}+P_i^{GW}$. These definitions correspond to $\alpha=1=\beta$ in the parametrization given in Eq. (3.18) of \cite{Compere:2019gft}. The Kerr black hole has $\mathcal E=M$ and $\mathcal J_z= M a$ with the standard orientation $\epsilon_{tr\theta\phi}=1$.} 
\begin{subequations}\label{fluxmultipole}
\begin{align}
	\Pi_0 &= 4\sum^{+\infty}_{\ell=2} 
	\biggl\{ \frac{(\ell+1)(\ell+2)}{(\ell-1)\ell \ell!(2\ell+1)!!}
	\!\!\stackrel{(\ell+1)}{M}_{\!\!\!L} \stackrel{(\ell+1)}{M}_{\!\!\!L}
	+ \frac{4\ell
		(\ell+2)}{(\ell-1)(\ell+1)!(2\ell+1)!!}
	\!\!\stackrel{(\ell+1)}{S}_{\!\!\!\! L}
	\stackrel{(\ell+1)}{S}_{\!\!\!\! L}\biggr\} \,,\label{FluxE}\\ 
%%%%%%%%%%%%%%%%%%%%%%%%%%%%%%%%%%%%%%%%%%%%%%%%%%%
	\Pi_i &=
	12\sum^{+\infty}_{\ell=2}  \biggl\{
	\frac{2(\ell+2)(\ell+3)}{\ell(\ell+1)!(2\ell+3)!!}
	\!\!\stackrel{(\ell+2)}{M}_{\!\!\!iL} \stackrel{(\ell+1)}{M}_{\!\!\!L} + \frac{8(\ell+3)}{(\ell+1)!(2\ell+3)!!}
	\!\!\stackrel{(\ell+2)}{S}_{\!\!\!\!iL}
	\!\stackrel{(\ell+1)}{S}_{\!\!\!\!L}\nn\\ & \qquad\qquad\qquad+ \frac{8(\ell+2)}{(\ell-1)(\ell+1)!(2\ell+1)!!}  \,\eps_{ijk}
	\!\!\stackrel{(\ell+1)}{M}_{\!\!\!jL-1}
	\!\stackrel{(\ell+1)}{S}_{\!\!\!\!kL-1}\biggr\} \,,\label{FluxP}\\
%%%%%%%%%%%%%%%%%%%%%%%%%%%%%%%%%%%%%%%%%%%%%%%%%%%%%%%%%
	\Sigma_i &= \eps_{ijk} \sum^{+\infty}_{\ell=2}
	 \biggl\{
	\frac{(\ell+1)(\ell+2)}{(\ell-1)\ell!(2\ell+1)!!}
	\!\stackrel{(\ell)}{M}_{\!jL-1} \!\stackrel{(\ell+1)}{M}_{\!\!\!kL-1}
	+ \frac{4\ell^2
		(\ell+2)}{(\ell-1)(\ell+1)!(2\ell+1)!!}
	\!\stackrel{(\ell)}{S}_{\!jL-1}
	\!\stackrel{(\ell+1)}{S}_{\!\!\!\!kL-1}\biggr\} \,,\label{FluxJ}\\ 
%%%%%%%%%%%%%%%%%%%%%%%%%%%%%%%%%%%%%%%%%%%%%%%%%%%%%%%%
\Lambda_i &=
	\sum^{+\infty}_{\ell=2} 
	\biggl\{\frac{(\ell+2)(\ell+3)}{\ell\,\ell!(2\ell+3)!!}\! \Bigl(
	\stackrel{(\ell+1)}{M}_{\!\!\!iL} \stackrel{(\ell+1)}{M}_{\!\!\!L} - \stackrel{(\ell)}{M}_{\!L} \stackrel{(\ell+2)}{M}_{\!\!\!iL}\Bigr)+
	\frac{4(\ell+3)}{\ell!(2\ell+3)!!}\!
	\Bigl( \stackrel{(\ell+1)}{S}_{\!\!\!\!iL}
	\stackrel{(\ell+1)}{S}_{\!\!\!\!L} - \stackrel{(\ell)}{S}_{\!L}
	\stackrel{(\ell+2)}{S}_{\!\!\!\!iL}\Bigr) \biggr\}  \,.\label{FluxG}
	\end{align}\end{subequations}
%
%%%%%%%%%%%%%%%%%%%%%%%%%%%%%%%%%%%%%%%%%%%%%%%%%%%%%%%%
The center-of-mass flux formula $\Lambda_i$ was derived in \cite{Compere:2019gft} and matches the formula obtained by Nichols in terms of spherical harmonics~\cite{Nichols:2017rqr} after erratum. It also reproduces the low harmonic results of~\cite{KQ16,KNQ18}. In~\cite{Blanchet:2018yqa}, the following simpler alternative center-of-mass flux was derived: 
\begin{align}
   \tilde\Lambda_i &=
	\sum^{+\infty}_{\ell=2} 
	\biggl\{\frac{2(\ell+2)(\ell+3)}{\ell\,\ell!(2\ell+3)!!}\!
	\stackrel{(\ell+1)}{M}_{\!\!\!iL} \stackrel{(\ell+1)}{M}_{\!\!\!L}+
	\frac{8(\ell+3)}{\ell!(2\ell+3)!!}\!
	\stackrel{(\ell+1)}{S}_{\!\!\!\!iL}
	\stackrel{(\ell+1)}{S}_{\!\!\!\!L}\biggr\} \,.\label{FluxGBF}  
\end{align}
It is associated with a different center-of-mass vector $\tilde G_i$. Since $\tilde \Lambda_i - \Lambda_i=\partial_{\tilde u} \Delta G_i$ is a total $\tilde u$ derivative, which vanishes on average for periodic systems, the two formulations of the flux-balance law are physically equivalent. One can relate the corresponding center-of-mass vectors by $\tilde G_i = G_i +\Delta G_i$ and the flux-formulae differ due to this shift at fixed retarded time $\tilde u$. A disadvantage of the shifted $\tilde G_i$, as was shown in~\cite{Compere:2019gft}, is that it cannot be written by means of a covariant formula over the sphere in Bondi gauge. We will use the definition $G_i$ and the corresponding flux $\Lambda_i$ henceforth. 

Notice that one only needs the leading term~\eqref{eq:zmunu} in the $1/\tl r$ expansion of the linearized metric in order to compute the energy and linear momentum fluxes, whereas in the case of the angular momentum and center-of-mass fluxes, one also needs to include the next-to-leading correction $1/\tl r^2$ (see~\cite{Blanchet:2018yqa} for details). On the other hand, the Poincar\'e fluxes become exact expressions when rewritten in terms of radiative multipole moments $U_L= {M}_{L}^{(\ell)}+\mathcal{O}(G)$, $V_L = {S}_{L}^{(\ell)}+\mathcal{O}(G)$~\cite{Compere:2019gft}.

\subsection{Quadratic memory in Newman-Unti coordinates}

Given a metric perturbation $h^\mn = G\,h_{1}^\mn + G^2\,h_{2}^\mn + \mathcal{O}(G^3)$ in an arbitrary (not necessarily harmonic) coordinate system $\{\tl u, \tl r, \tl\theta^a\}$, we constructed perturbatively  in~\cite{Blanchet:2020ngx} the gauge transformation $\{\tl u, \tl r, \tl\theta^a\}\longrightarrow\{u,r,\theta^a\}$ required to reach the Newman-Unti (NU) gauge. Denoting the coordinate transformation as
\begin{subequations}\label{eq:transfBondi}
	\begin{align}
	u &= \tl u + G U_{1}(\tl u,\tl r, \tl \theta^a) + G^2 U_{2}(\tl u,\tl r, \tl \theta^a) + \mathcal{O}(G^3)\,,\\[0.15cm]
	r &= \tl r + G R_{1}(\tl u,\tl r, \tl \theta^a) + G^2 R_{2}(\tl u,\tl r, \tl \theta^a) + \mathcal{O}(G^3) \,,\\[0.15cm]
	\theta^a &= \tl \theta^a + G \Theta^a_{1}(\tl u,\tl r, \tl \theta^a) + G^2 \Theta^a_{2}(\tl u,\tl r, \tl \theta^a) + \mathcal{O}(G^3)\,,
	\end{align}
\end{subequations}
For the transformed metric to satisfy NU gauge $g_{rr}=0=g_{ra}$ and $g_{ur}=-1$, we have to solve successively the linear order equations (where $\tl k^\mu$ is the Minkowski null outgoing vector defined in the original coordinates)
\begin{subequations}\label{eq:linorder}
	\begin{align}
	\tl k^{\mu}\tl \partial_\mu U_1 &= \frac{1}{2}\tl k_{\mu}\tl k_{\nu}h_1^{\mu\nu}\,,\label{eq:eqU1}\\
	\tl k^{\mu}\tl \partial_\mu R_1 &= \frac{1}{2} h_1 + \tl n_i\left( \tl \p_i U_1 -\tl k_\m  h_{1}^{\mu i} \right)\,,\label{eq:eqR1}\\
	\tl k^{\mu}\tl \partial_\mu \Theta^a_1 &= \frac{\tl e^a_i}{\tl r}\left(\tl \p_i U_1 - \tl k_\mu h_1^{\mu i} \right) \,,\label{eq:eqTheta1}
	\end{align}
\end{subequations}
and next, the quadratic order equations
\begin{subequations}\label{eq:quadorder}
\begin{align}
\tl k^{\mu}\tl \partial_\mu U_2 &= \frac{1}{2}\tl k_{\mu}\tl k_{\nu}h_2^{\mu\nu} + \left(\frac{1}{2}\tl\partial^{\mu}U_1 - \tl k_\nu h_1^{\mu\nu}\right)\tl\partial_{\mu}U_1\,,\label{eq:eqU2}\\
\tl k^{\mu}\tl \partial_\mu R_2 &= \frac{1}{8}h_1^2 -\frac{1}{4} h_1^{\mu\nu}h_{1 \mu\nu} + \frac{1}{2} h_2 + \tl n_i\left( \tl \p_i U_2 -\tl k_\m  h_{2}^{\mu i} + (\tl \p_\m U_1)h_1^{\m i} \right)\nn\\ & \quad+ \left(\tl\partial^{\mu}U_1 - \tl k_\nu h_1^{\mu\nu}\right)\tl\partial_{\mu}R_1\,,\label{eq:eqR2}\\
\tl k^{\mu}\tl \partial_\mu \Theta^a_2 &= \frac{\tl e^a_i}{\tl r}\left(\tl \p_i U_2 - \tl k_\mu h_2^{\mu i}+(\tl \p_\m U_1)h_1^{\mu i} \right) + \left(\tl\partial^{\mu}U_1 - \tl k_\nu h_1^{\mu\nu}\right)\tl\partial_{\mu}\Theta^a_1\,.\label{eq:eqTheta2}
\end{align}
\end{subequations}
In this section, we compute the memory-type and mass-loss-type hereditary terms in the NU metric, following the previous algorithm. The tail-type hereditary terms (in the quadrupole case) have been investigated in~\cite{Blanchet:2020ngx} and we will only report the result. Note that the first order perturbation $h_1^\mn$ does not contain hereditary integrals (it is instantaneous in terms of the canonical moments $M_L$ and $S_L$). Only the second order metric $h_2^\mn$ involves the mass-loss and memory terms. Moreover, the differential operator $\tilde k^\mu \tilde{\partial}_{\mu}={\tilde \partial}_{r}\vert_{\tilde u}$ is simply solved by integration over $\tilde{r}$ at constant $\tilde{u}$. Therefore, no hereditary integrals can be generated from instantaneous (non-hereditary) terms on the right-hand side of the equations~\eqref{eq:quadorder} and, in order to control them, it is sufficient to solve the linear equations 
\begin{subequations}\label{eq:quadordersim}
\begin{align}
\tl k^{\mu}\tl \partial_\mu U_2 &= \frac{1}{2}\tl k_{\mu}\tl k_{\nu}h_2^{\mu\nu}\,,\label{eq:quadordersima} \\
\tl k^{\mu}\tl \partial_\mu R_2 &= \frac{1}{2} h_2 + \tl n_i\left( \tl \p_i U_2 -\tl k_\m  h_{2}^{\mu i} \right)\,,\\
\tl k^{\mu}\tl \partial_\mu \Theta^a_2 &= \frac{\tl e^a_i}{\tl r}\left(\tl \p_i U_2 - \tl k_\mu h_2^{\mu i} \right)\,.
\end{align}
\end{subequations}

We now implement our algorithm by solving the system~\eqref{eq:quadordersim}, focusing on the memory and mass-loss terms. Reminding the property $\tl k_\nu K^\mn = 0$, we see from eq.~\eqref{eq:h2mem} that the first equation~\eqref{eq:quadordersima} reduces to $\tl k^{\mu}\tl \partial_\mu U_2 = \mathcal{O}(\tilde r^{-2})$ at leading order, which is directly solved to $U_2=U^0_2(\tl u,\tl{\bm{n}}) + \mathcal{O}(\tilde r^{-1})$. After imposing asymptotic flatness and setting the BMS transformation at second order to zero, we get $U^0_2(\tl u,\tl{\bm{n}})=0$. Therefore, the only contributions to $U_2$ come from the $1/\tl r^2$ term in $h_2^\mn$ as given by eq.~\eqref{eq:h2mem}. This term is easily integrated and we obtain
\begin{align}\label{eq:U2}
U_2{\Big|}_\text{mem} = \frac{\tilde n_i}{\tilde r} \left(-\frac{1}{6}\stackrel{(-2)}{\Pi_i} + \,2 \!\stackrel{(-1)}{\Lambda_i}\right)\,.
\end{align}
We recall that $\Pi_i$ is the flux of linear momentum while $\Lambda_i$ is the flux of center-of-mass [see eqs.~\eqref{fluxes}--\eqref{fluxmultipole}], all quantities being evaluated at $\tilde u\equiv\tilde t-\tilde r$ and $\tl x^{a}$. Continuing the algorithm, we find that $R_2$ does not contain memory/mass-loss terms whereas $\Theta^{a}_2$ receives contributions from $\Pi_i$, $\Lambda_i$, as well as the angular momentum flux $\Sigma_i$:
\begin{subequations}\label{eq:R2Th2}
	\begin{align}
	R_2{\Big|}_\text{mem} &=0\,,\\
	\Theta^{a}_2{\Big|}_\text{mem} &=\frac{e^{a}_i}{\tilde r^2} \left(\frac{1}{12}\stackrel{(-2)}{\Pi_i} +  \stackrel{(-1)}{\Lambda_i} +\,\eps_{ipq}\tilde n_p \stackrel{(-1)}{\Sigma_q} \right)\,.
	\end{align}
\end{subequations}
We refer to the previous paper~\cite{Blanchet:2020ngx} for the computation of the tail contributions due to mass quadrupole in the quantities~\eqref{eq:U2}--\eqref{eq:R2Th2}.

We proceed with the next steps as described in~\cite{Blanchet:2020ngx}: we successively compute the contravariant components of the metric, $g^{rr}$, $g^{ra}$ and $g^{ab}$, and deduce its covariant components in the NU gauge as $g_{ua}=g^{rb}g_{ab}$, $g_{uu}=-g^{rr}+g^{ra}g_{ua}$ and $g_{ab}=(g^{ab})^{-1}$. In the end, we re-express the metric in terms of the NU coordinates $\{u,r,\theta^a\}$ using the inverse of eqs.~\eqref{eq:transfBondi}. In particular, this entails re-expanding the 2-sphere metric as 
\begin{equation}\label{eq:gamab}
{\tl r^2}\tl \gamma_{ab} =   r^2 \Bigl[\gamma_{ab} - 2G^2\Bigl( r^{-1} R_{2} \gamma_{ab} + \Theta_2^c  \Gamma^e_{c(a} \gamma_{b)e}\Bigr) \Bigr] + \mathcal{O}(G^3)\,,
\end{equation} 
where $ \Gamma^a_{bc}$ is the Christoffel symbol associated with the covariant derivative on the sphere. This brings a memory correction coming from $\Theta_2^a$ by virtue of~\eqref{eq:R2Th2}. We find that the Christoffel symbols cancel out so that our final metric is covariant with respect to diffeomorphisms acting on the sphere. We finally get
\begin{subequations}\label{eq:gmunu}
\begin{align}
g_{uu}{\Big|}_\text{mem} &= -1 - G^2\biggl[\frac{1}{r}\biggl(\frac{1}{2}\stackrel{(-1)}{\Pi_0}   + \frac{1}{2}n_i\stackrel{(-1)}{\Pi_i}\biggr) + \frac{n_i}{r^2}\biggl(\frac{1}{6}\stackrel{(-2)}{\Pi_i} + 2 \stackrel{(-1)}{\Lambda_i}\biggr)\biggr] + \mathcal{O}(G^3)\,,\\
g_{ua}{\Big|}_\text{mem} &=  -G^2 e^i_{a} \biggl[ \frac{1}{2} \sum_{\ell\geqslant 2}\frac{1}{\ell +1} n_{L-1} \stackrel{(-1)}{\Pi_{iL-1}} + \frac{1}{ r}\biggl( \frac{1}{6}\stackrel{(-2)}{\Pi_i} + 2 \stackrel{(-1)}{\Lambda_i} + 2\eps_{ipq} \,n_p \stackrel{(-1)}{\Sigma_q}\biggr)\biggr] +\mathcal{O}(G^3)\,,\\
%g_{ua}{\Big|}_\text{mem} &=  -G^2 e^i_{a} \biggl[ \frac{1}{2} \sum_{\ell\geqslant 2}\frac{1}{\ell +1} n_{L-1} \stackrel{(-1)}{\Pi_{iL-1}} \nn\\&\quad\quad\quad\quad + \frac{1}{ r}\biggl( \frac{1}{6}\stackrel{(-2)}{\Pi_i} + 2 \stackrel{(-1)}{\Lambda_i} + 2\eps_{ipq} \,n_p \stackrel{(-1)}{\Sigma_q}\biggr)\biggr] +\mathcal{O}(G^3)\,,\\
g_{ab}{\Big|}_\text{mem} &= r^2 \gamma_{ab} + 2G^2\, r\,e_{\langle a}^i e_{b\rangle}^j\sum_{\ell\geqslant 2}\frac{1}{(\ell +1)(\ell +2)} n_{L-2} \!\!\stackrel{(-1)}{\Pi}\!\!{}_{ijL-2} + \mathcal{O}(G^3)\,.
\end{align}
\end{subequations}
The above metric is entirely expressed in terms of the NU coordinates $\{u,r,\theta^a\}$. Now, the general asymptotically flat solution of interest to the vacuum Einstein's equations in Newman-Unti coordinates, up to $\mathcal{O}(r^{-3})$ corrections, reads \cite{newman1963class,Barnich:2011ty,Barnich:2011mi,Flanagan:2015pxa,Barnich:2019vzx,Blanchet:2020ngx,Seraj:2021rxd}
\begin{subequations}\label{eq:gmunuBondiform}
\begin{align}
	g_{uu} &= -1+\frac{2 \left(m+\frac{1}{8}C_{ab}N^{ab}\right)}{r}-\frac{D_a N^a}{3r^2}+\mathcal{O}(r^{-3})\,, \label{eq:guuBondiform}\\
	g_{ua} &= \frac{D^b C_{ab}}{2}+\frac{2N_a}{3r}+\mathcal{O}(r^{-2})\,, \label{eq:guaBondiform}\\ g_{ab} &= r^2 \gamma_{ab}+r \,C_{ab} + \frac{C^{cd}C_{cd}}{8}\gamma_{ab}+\mathcal{O}(r^{-1})\,,\label{eq:gabBondiform} 
\end{align}
\end{subequations}
together with $g_{rr}=g_{ra}=0$ and $g_{ur}=-1$. Here, $m$ is the Bondi mass aspect, $C_{ab}$ is the shear,\footnote{The shear $C_{ab}$ may be set traceless by a suitable choice of origin for the radial coordinate \cite{Barnich:2011ty}.} $N_{ab}$ is the news and $N_a$ is the angular momentum aspect.

We can thus immediately read off the memory terms in the Bondi mass aspect $m$, the angular momentum aspect $N_a$ and the Bondi shear $C_{ab}$, as
\begin{subequations}\label{eq:mNaCab}
\begin{align}
m{\Big|}_\text{mem} &= -G^2\left(  \frac{1}{4}\stackrel{(-1)}{\Pi_0} + \frac{1}{4}n^i\!\stackrel{(-1)}{\Pi_i}\right) + \mathcal{O}(G^3)\,,\\
N_{a}{\Big|}_\text{mem} &= -G^2 e^i_{a} \left(  \frac{1}{4}\!\stackrel{(-2)}{\Pi_i} +3 \!\stackrel{(-1)}{\Lambda_i} + 3\eps_{ipq} \,n_p \!\stackrel{(-1)}{\Sigma_q}\right)+ \mathcal{O}(G^3)\,,\\ \label{eq:Cabmem}
C_{ab}{\Big|}_\text{mem} &= 2G^2 e_{\langle a}^i e_{b\rangle}^j\sum_{\ell\geqslant 2}\frac{1}{(\ell +1)(\ell +2)} \,n_{L-2} \!\!\stackrel{(-1)}{\Pi}\!\!{}_{ijL-2}+ \mathcal{O}(G^3)\,.
\end{align}
\end{subequations}
There is no memory contribution in the combination $\dot{N}_{a} - D_a m$, consistently with one of the Einstein field equations. The shear~\eqref{eq:Cabmem} is actually traceless and agrees with eq.~(5.6) of \cite{Blanchet:1997ji} after expressing the linearized gothic metric as minus the standard linearized metric. In fact, comparing the metric \eqref{eq:gmunuBondiform} in Newman-Unti coordinates up to $\mathcal{O}(r^{-3})$ with our results~\eqref{eq:gmunu}, we find that they are perfectly consistent. Namely, we obtain from eq.~\eqref{eq:mNaCab} the divergences
\begin{subequations}
	\begin{align}
	D_a N^{a}{\Big|}_\text{mem} &= G^2 n^i \left( \frac{1}{2}\stackrel{(-2)}{\Pi_i} + 6 \stackrel{(-1)}{\Lambda_i}\right)+ \mathcal{O}(G^3)\,,\\ 
	D^b C_{ab}{\Big|}_\text{mem} &= -G^2 e_{a}^i \sum_{\ell\geqslant 2}\frac{1}{\ell +1} \,n_{L-1} \!\!\!\stackrel{(-1)}{\Pi}\!\!\!{}_{iL-1} + \mathcal{O}(G^3)\,, 
	\end{align}
\end{subequations}
and we see that the first expression agrees with our result for the sub-dominant term $1/r^2$ in $g_{uu}$ while the second one agrees with twice the $r^0$ term in $g_{ua}$. 

We can also express the NU metric corresponding to the memory terms of the 
quadrupole-quadrupole interaction $M_{ij} \times M_{ij}$ in terms of the canonical mass quadrupole $M_{ij}$.  The corresponding quadrupole-quadrupole metric in harmonic coordinates was computed in~\cite{Blanchet:1997ji}. From eqs.~(4.5) of~\cite{Blanchet:1997ji}, for quadratic products of two mass quadrupoles, the only non-zero multipolar coefficients are 
\begin{align}\label{eq:Piquad}
\Pi_0 = \frac{4}{5} \stackrel{(3)}{M}\!{}_{ij}\stackrel{(3)}{M}\!{}_{ij}\,,\qquad\Pi_{ij} = -\frac{24}{7} \stackrel{(3)}{M}\!{}_{k \langle i}\stackrel{(3)}{M}\!{}_{j \rangle k}\,,\qquad\Pi_{ijkl} = \stackrel{(3)}{M}\!{}_{\langle ij}\stackrel{(3)}{M}\!{}_{kl\rangle}\,.
\end{align}
This yields
\begin{subequations}\label{eq:gmunuquadmem}
	\begin{align}
	g_{uu}{\Big|}_\text{mem} &= -1 - \frac{2G^2}{5r}\!\int_{-\infty}^{u} \dd v\stackrel{(3)}{M}\!{}_{ij}(v)\stackrel{(3)}{M}\!{}_{ij}(v)\,,\\
	g_{ua}{\Big|}_\text{mem} &=  G^2 e^i_{a} n^j \int_{-\infty}^{u} \dd v \biggl[ \frac{9}{8} \stackrel{(3)}{M}\!{}_{k \langle i}\stackrel{(3)}{M}\!{}_{j \rangle k}
	-\frac{4}{5r}\left(\stackrel{(2)}{M}\!{}_{i k}\stackrel{(3)}{M}\!{}_{jk} - \stackrel{(2)}{M}\!{}_{j k}\stackrel{(3)}{M}\!{}_{ik}\right)\biggl](v)\,,\\
	g_{ab}{\Big|}_\text{mem} &= r^2 \left[\gamma_{ab} + \frac{G^2}{r}\,e_{\langle a}^i e_{b\rangle}^j\int_{-\infty}^{u} \dd v\left(-\frac{4}{7}\stackrel{(3)}{M}\!{}_{ki}\stackrel{(3)}{M}\!{}_{jk} + \frac{1}{15}n_{kl}\stackrel{(3)}{M}\!{}_{\langle ij}\stackrel{(3)}{M}\!{}_{kl\rangle}\right)(v)\right] \,.
	\end{align}
\end{subequations}

In terms of physical fluxes~\eqref{fluxmultipole}, the memory contributions to the Bondi mass and angular momentum aspects take the simple and explicit form 
\begin{subequations}
\begin{align}
    m{\Big|}_\text{mem} &= -G\bigl(E^{\text{GW}}+3n_i P_i^{\text{GW}}\bigr)+\mathcal{O}(G^3)\,, \\
    N_a{\Big|}_\text{mem} &= -3Ge^i_a\bigl(G_i^{\text{GW}}+\eps_{ipq}n_p S_q^{\text{GW}}\bigr)+\mathcal{O}(G^3)\,. 
\end{align}
\end{subequations}

Let us also derive, to end with, an interesting alternative expression for the shear, which is related to the asymptotic waveform by $C_{ab}=e_{\langle a}^i e_{b\rangle}^j H^\text{TT}_{ij}$ [see \emph{e.g.},~\cite{Blanchet:2020ngx} and eq.~\eqref{eq:TTWF} above]. Using eq.~\eqref{eq:energidensity} and eq.~\eqref{eq:PiSTF}, we obtain 
\begin{align}\label{eq:shear0}
C_{ab}{\Big|}_\text{mem} = 8 e_{\langle a}^i e_{b\rangle}^j \sum_{\ell \geqslant 2}\frac{(2\ell+1)!!}{(\ell+2)!}\,n_{L-2} \int \dd\Omega'\,\hat{n}'_{ijL-2}\,\frac{\dd E^\text{GW}}{\dd\Omega'}(u,\bm{n}')\,,
\end{align}
where the infinite multipole series therein can be summed up in closed form as (see below for the proof)
\begin{align}\label{eq:closedform}
e_{\langle a}^i e_{b\rangle}^j\,\sum_{\ell \geqslant 2}\frac{(2\ell+1)!!}{(\ell+2)!}\,n_{L-2}\,\hat{n}'_{ijL-2} = \frac{1}{2} e_{\langle a}^i e_{b\rangle}^j \,\frac{n'_{i}n'_{j}}{1-\bm{n}\cdot\bm{n}'}\,.
\end{align}
This leads to the following elegant result, which constitutes the best interpretation of the non-linear memory effect as due to the re-radiation of GWs by gravitons~\cite{1987Natur.327..123B,Christodoulou:1991cr,Th92}:\footnote{The derivation of the result~\eqref{eq:shear1} in~\cite{1987Natur.327..123B,Christodoulou:1991cr,Th92} is very different from the one adopted here, which is rather based on~\cite{Blanchet:1992br,Wiseman:1991ss}.}
\begin{align}\label{eq:shear1}
C_{ab}{\Big|}_\text{mem} = 4 e_{\langle a}^i e_{b\rangle}^j \int \dd\Omega'\,\frac{n'_{i}n'_{j}}{1-\bm{n}\cdot\bm{n}'}\,\frac{\dd E^\text{GW}}{\dd\Omega'}(u,\bm{n}')\,,
\end{align}
where the factor $1/(1-\bm{n}\cdot\bm{n}')$ is reminiscent of the Li\'enard-Wiechert potentials in the case of massless gravitons. It exactly reproduces the result of Christodoulou \cite{Christodoulou:1991cr} and Thorne~\cite{Th92} (see also the earlier results \cite{PhysRevD.28.1894,1989GReGr..21.1205L}).

\paragraph{Proof of the formula \texorpdfstring{\eqref{eq:closedform}}{}.} We consider the following TT projection with respect to the unit vector $\bm{n}=(n_i)$:
\begin{align}\label{eq:proof1}
	A^\text{TT}_{ij} \equiv \biggl[\frac{n'_{i}n'_{j}}{1-\bm{n}\cdot\bm{n}'}\biggr]^\text{TT} \equiv \,\, \perp_{ijkl}^\text{TT} \frac{n'_{k}n'_{l}}{1-\bm{n}\cdot\bm{n}'}\,,
\end{align}
where we recall that the TT projection reads $\perp_{ijkl}^\text{TT} = \frac{1}{2}(\perp_{ik}\perp_{jl} + \perp_{jk}\perp_{il} - \perp_{ij}\perp_{kl})$, with the usual perpendicular operator $\perp_{ij}=\delta_{ij}-n_i n_j$. We expand the denominator in eq.~\eqref{eq:proof1} as a power series in $\bm{n}\cdot\bm{n}'$ (which is convergent as soon as $\bm{n}\cdot\bm{n}'<1$), hence
\begin{align}\label{eq:proof2}
	A^\text{TT}_{ij} = \biggl[ \sum_{\ell=2}^{+\infty} n_{L-2} \,n'_{ijL-2}\biggr]^\text{TT}\,.
\end{align}
The multi-index $L-2$ contains $\ell-2$ indices, namely $a_1\cdots a_{\ell-2}$. Using eq.~(A.21a) in~\cite{1986RSPTA.320..379B}, we transform the ordinary product of unit vectors $n'_{ijL-2}$ into a sum of STF products
\begin{subequations}\label{eq:proof3}
\begin{align}
	A^\text{TT}_{ij} &= \biggl[ \sum_{\ell=2}^{+\infty} n_{L-2} \sum_{k=0}^{[\frac{\ell}{2}]} \alpha_k^\ell \,\delta_{\{2K}\,\hat{n}'_{L-2K\}}\biggr]^\text{TT}\,,\label{eq:proof3a}\\
	\text{with}\quad \alpha_k^\ell &\equiv \frac{(2\ell-4k+1)!!}{(2\ell-2k+1)!!}\,.\label{eq:proof3b}
\end{align}
\end{subequations}
The operation over indices $\{\}$ is defined as the un-normalized sum over the smallest set of permutations of $i_1\cdots i_\ell$ which makes the object symmetrical in $L=i_1\cdots i_\ell$. The object $\delta_{\{2K}\,\hat{n}'_{L-2K\}}$ contains $\ell$ indices $L=ijL-2$ (with, say, $i = a_\ell$ and $j = a_{\ell-1}$) of which $2k$ are displayed onto the product of $k$ Kronecker symbols denoted $\delta_{2K}$. As an example we have $\delta_{\{ab}n_{c\}} \equiv \delta_{ab}n_c + \delta_{bc}n_a + \delta_{ca}n_b$. 

The point is that, among all the terms composing $\delta_{\{2K}\,\hat{n}'_{L-2K\}}$, we can discard all those which contain either $\delta_{ij}$, $\delta_{i a_p}$ or $\delta_{j a_q}$, since such terms will be cancelled by the TT projection. This is obvious for $\delta_{ij}$; in the two other cases, this results from the fact that, after multiplication by $n_{L-2}$ in eq.~\eqref{eq:proof3a}, $\delta_{i a_p}$ or $\delta_{j a_q}$ will yield some $n_i$ or $n_j$. So, we can rewrite the expression~\eqref{eq:proof3a} by excluding the indices $ij$ from the operation $\{\}$, which we indicate by underlining the two indices $ij$: 
\begin{align}\label{eq:proof4}
	A^\text{TT}_{ij} = \biggl[ \sum_{\ell=2}^{+\infty} n_{L-2} \sum_{k=0}^{[\frac{\ell-2}{2}]} \alpha_k^\ell \,\delta_{\{2K}\,\hat{n}'_{\underline{ij}L-2-2K\}}\biggr]^\text{TT}\,.
\end{align}
The next step is to notice from eq.~(A.19) in~\cite{1986RSPTA.320..379B} that the number of terms composing the object $\delta_{\{2K}\,\hat{n}'_{\underline{ij}L-2-2K\}}$ is $\frac{(\ell-2)!}{2^k k!(\ell-2-2k)!}$. When contracted with $n_{L-2}$, all these terms will merge into a single one for each values of $\ell$ and $k$. Thus, we have 
\begin{align}\label{eq:proof5a}
	A^\text{TT}_{ij} = \Biggl[ \sum_{\ell=2}^{+\infty} \sum_{k=0}^{[\frac{\ell-2}{2}]} \alpha_k^\ell \,\frac{(\ell-2)!}{2^k k!(\ell-2-2k)!}\, n_{L-2-2K}\,\hat{n}'_{ijL-2-2K}\Biggr]^\text{TT}\,.
\end{align}
We change $\ell$ into $\ell+2k$ and rewrite the previous expression as
\begin{align}\label{eq:proof5b}
	A^\text{TT}_{ij} = \Biggl[ \sum_{\ell=2}^{+\infty} \sum_{k=0}^{+\infty} \alpha_k^{\ell+2k} \,\frac{(\ell+2k-2)!}{2^k k!(\ell-2)!}\, n_{L-2}\,\hat{n}'_{ijL-2}\Biggr]^\text{TT} = \Biggl[ \sum_{\ell=2}^{+\infty} \,S_\ell \,n_{L-2}\,\hat{n}'_{ijL-2}\Biggr]^\text{TT}\,,
\end{align}
introducing the coefficient $S_\ell$ which is given as an infinite series over all integer values of $k$. However, we find, using the expression of the coefficients $\alpha_k^{\ell+2k}$ deduced from eq.~\eqref{eq:proof3b}, that this series can actually be re-summed in closed analytic form:
\begin{align}\label{eq:proof5c}
	S_\ell = \frac{(2\ell+1)!!}{(\ell-2)!} \sum_{k=0}^{+\infty} \,\frac{(\ell+2k-2)!}{2^k k!(2\ell+2k+1)!!} = 2^{\ell+2}\frac{\Gamma\bigl(\ell+\frac{3}{2}\bigr)}{\sqrt{\pi}\,\Gamma(\ell+3)} = 2\frac{(2\ell+1)!!}{(\ell+2)!}\,.
\end{align}
Hence we obtain the simple result
\begin{align}\label{eq:proof5d}
	A^\text{TT}_{ij} = \Biggl[ 2\sum_{\ell=2}^{+\infty} \frac{(2\ell+1)!!}{(\ell+2)!}\, n_{L-2}\,\hat{n}'_{ijL-2}\Biggr]^\text{TT}\,.
\end{align}
Recalling that $e_{\langle a}^i e_{b\rangle}^j\perp_{ijkl}^\text{TT} \,= e_{\langle a}^k e_{b\rangle}^l$, we have therefore proved the formula~\eqref{eq:closedform}.

\subsection{\texorpdfstring{$M_{ij} \times M_{ij}$}{} and \texorpdfstring{$M \times M_{ij}$}{} asymptotic data in NU gauge}

%\Geo{Added:} 
We now obtain the complete NU metric corresponding to the quadrupole-quadrupole interaction $M_{ij} \times M_{ij}$ and tails $M \times M_{ij}$, \emph{i.e.}, including, besides the non-linear memory and mass-loss terms derived in the previous section and besides the tail terms obtained previously in~\cite{Blanchet:2020ngx}, all the instantaneous (non-hereditary) terms.

In this complete calculation of the $M_{ij} \times M_{ij}$ interaction, we start from the metric in harmonic coordinates as given in~\cite{Blanchet:1997ji}, instead of the modified harmonic coordinates described in Sec~\ref{sec:modharmcoord}, in which the metric concerning hereditary effects was already free of far-zone logarithms [see eq.~\eqref{eq:h2mem}]. Applying our algorithm, we shall check that indeed all the hereditary terms besides the memory, and all associated far-zone logarithms, disappear in the end of the calculation. A more sophisticated treatment of hereditary terms, which is explained in the Appendix~\ref{app:A} below, is however required. We shall of course recover in particular the memory terms computed in the previous section.\footnote{Our practical calculation is done with the software \emph{Mathematica} supplemented by the \emph{xAct} package~\cite{xtensor}.}

Our final results for the NU metric (see Sec.~\ref{sec:NUBondi} for the link with the Bondi metric) read 
\begin{subequations}\label{eq:gmunuNUform}
	\begin{align}
		g_{uu} &= -1+\frac{2\bigl(m+\frac{1}{8}C_{ab}N^{ab}\bigr)
		}{r}-\frac{D_a N^a}{3r^2}+\sum_{n=3}^{6}\frac{1}{r^{n}}\!\stackrel[(n)]{}{g_{uu}}\,,\label{eq:guuNUform}\\
		g_{ua} &= \frac{D^b C_{ab}}{2}+\frac{2N_a}{3r}+\sum_{n=2}^{5}\frac{1}{r^{n}}\!\stackrel[(n)]{}{g_{ua}}\,,\label{eq:guaNUform}\\
		g_{ab} &= r^2\left[\left(1 + \frac{W}{r^2} \right) \gamma_{ab} + \frac{1}{r} e^{i}{}_{\langle a} e^{j}{}_{b\rangle}\biggl(H^{\mathrm{TT}}_{i j}+\sum_{n=2}^{+\infty}\frac{1}{r^{n}}\!\stackrel[(n)]{}{E_{ij}}\biggr)\right]\,.\label{eq:gabNUform}
	\end{align}
\end{subequations}
Here, $H^{\mathrm{TT}}_{i j}$ is the radiation field, which is a free data relating to the Bondi shear as $C_{a b}=e_{\langle a}^i e_{b\rangle}^j H^{\mathrm{TT}}_{i j}$. It can be decomposed into $\ell\geqslant 2$ mass/electric and current/magnetic radiative multipoles $U_L,V_L$ as 
\begin{align}\label{HijTT}
	H_{i j}^{\mathrm{TT}}&=4 \perp_{ijkl}^\mathrm{TT}\!(\bm{n}) \sum_{\ell=2}^{+\infty} \frac{1}{\ell !}\left\{\hat n_{L-2} \mathrm{U}_{k l L-2}(u)-\frac{2 \ell}{%c
		\ell+1} \hat n_{p L-2} \eps_{p q(k} \mathrm{V}_{l) q L-2}(u)\right\}\,.
\end{align}
The nonzero radiative multipole moments are given by (displaying only the terms that are linear in $M_{ij}$ or that correspond to the $M_{ij} \times M_{ij}$ and $M \times M_{ij}$ interactions)\footnote{For more general expressions involving current moments and higher multipoles of mass moments, see Sec.~3.3 of \cite{Blanchet:2013haa} and references therein.} 
\begin{subequations}\label{HijTTexp}
	\begin{align}
		U_{ij} &= G M^{(2)}_{ij}+2G^2 M \int_{0}^{+\infty} \dd z \left[ \ln \left(\frac{z}{2b_0 }\right)+\frac{11}{12}\right]{M}_{ij}^{(4)}(u-z) \nn\\
        & + G^2 \left( -  \frac{2}{7} \int_{-\infty}^{u} \!\dd v M^{(3)}_{k\langle i}(v) M^{(3)}_{j\rangle k}(v) - \frac{2}{7} M^{(2)}_{k\langle i} M^{(3)}_{j\rangle k}
		-  \frac{5}{7} M^{(1)}_{k\langle i} M^{(4)}_{j\rangle k}
		+ \frac{1}{7} M_{k\langle i} M^{(5)}_{j\rangle k}\right)\,, \label{Uijtail}\\
		U_{ijkl} &= G^2\left(\frac{2}{5} \int_{-\infty}^{u} \!\dd v M^{(3)}_{\langle ij}(v) M^{(3)}_{kl\rangle}(v) - \frac{102}{5} M^{(2)}_{\langle ij} M^{(3)}_{kl\rangle}
		-  \frac{63}{5} M^{(1)}_{\langle ij} M^{(4)}_{kl\rangle}
		-  \frac{21}{5} M_{\langle ij} M^{(5)}_{kl\rangle}\right)\,,\\
		V_{ijk} &= G^2 \eps_{pq\langle i}\left(\frac{1}{2}  M^{(1)}_{j\underline{p}} M^{(4)}_{k\rangle q}
		-  \frac{1}{10} M_{j\underline{p}} M^{(5)}_{k\rangle q}\right)\,,
	\end{align}
\end{subequations}
where time derivatives of moments are denoted, \emph{e.g.}, $M^{(q)}_{ij}\equiv \stackrel{(q)}{M}\!{}_{ij}$, and underlined index must be regarded to be outside the STF projection. Here $b_0$ is a gauge constant related to the origin of time in radiative coordinates. 

The remaining data in NU gauge are the mass aspect $m$ and the angular momentum aspects $N_i=e_i{}^a N_a$, which read (adding also the leading linear terms) 
\begin{subequations} \label{eq:Bondiaspects}
	\begin{align}
		m &= G \left(M
		+ 3 {n}^{i} P_{i} 
		+ 3 \hat{n}^{ij} M^{(2)}_{ij}\right) \nn\\
		&
		+ G^2 \biggl[-  \frac{1}{5} \int_{-\infty}^{u} \!\dd v M^{(3)}_{ij}(v) M^{(3)}_{ij}(v) 
		+ \hat{n}^{ij} \biggl(- \frac{6}{7} M^{(2)}_{ik} M^{(3)}_{jk} -  \frac{15}{7} M^{(1)}_{ik} M^{(4)}_{jk} + \frac{3}{7} M_{ik} M^{(5)}_{jk}\biggr)\nn\\
		&\qquad~~ + \hat{n}^{ijkl} \biggl(- \frac{51}{4} M^{(2)}_{ij} M^{(3)}_{kl}
		-  \frac{63}{8} M^{(1)}_{ij} M^{(4)}_{kl} -  \frac{21}{8} M_{ij} M^{(5)}_{kl}\biggr)\biggl]\,,\label{mexpl}\\
		%%%%%%%%%%%%%%%%%%%%%%%%%%%%%%%%%%%%%%%%%%%%%%%%%%%%%%%%%%%%%%%%
		N_{i} &= 3G \left( M_{i}
		+ 2 {n}^{j} M^{(1)}_{ij}\right)^{\mathrm{T}} \nn\\
		&
		+ G^2 \biggl [ \frac{6}{5} {n}^{j} \int_{-\infty}^{u} \!\dd v \Bigl[M^{(2)}_{jk} M^{(3)}_{ik} - M^{(2)}_{ik} M^{(3)}_{jk}\Bigr](v) \nn\\
		&\qquad~~ - \frac{6}{35}{n}^{j}\Bigl( 14 M^{(2)}_{ik} M^{(2)}_{jk}
		+ 22 M^{(1)}_{jk} M^{(3)}_{ik}
		+ 22 M^{(1)}_{ik} M^{(3)}_{jk}
		+  M_{jk} M^{(4)}_{ik}
		+  M_{ik} M^{(4)}_{jk}\Bigr)\nn\\
		&\qquad~~ -  \frac{3}{2} \hat{n}^{jkl} \Bigl(9 M^{(2)}_{ij} M^{(2)}_{kl}
		+ 10 M^{(1)}_{jk} M^{(3)}_{il}
		+ 4 M^{(1)}_{ij} M^{(3)}_{kl}
		+ 3 M_{jk} M^{(4)}_{il} + 4 M_{ij} M^{(4)}_{kl}\Bigr)\biggl]^{\mathrm{T}}\,.\label{Naexpl}
	\end{align}
\end{subequations}
The trace of the angular part of the metric~\eqref{eq:gabNUform} is
\begin{align}
	W &= G^2\biggl[ \frac{1}{5} M^{(2)}_{ij} M^{(2)}_{ij}
	-  \frac{6}{7} \hat{n}^{ij} M^{(2)}_{ik} M^{(2)}_{jk}
	+ \frac{1}{4} \hat{n}^{ijkl} M^{(2)}_{ij} M^{(2)}_{kl}\nn\\
	&\qquad\qquad + \frac{1}{r^2} \biggl(\frac{3}{5} M_{ij} M^{(2)}_{ij} -  \frac{18}{7} \hat{n}^{ij} M_{ik}  M^{(2)}_{jk}
	+ \frac{3}{4} \hat{n}^{ijkl} M_{ij} M^{(2)}_{kl}\biggr)\nn\\
	&\qquad\qquad +\frac{1}{r^4} \biggl(\frac{7}{25} M_{ij} M_{ij}
	-  \frac{6}{5} \hat{n}^{ij} M_{ik} M_{jk} 
	+ \frac{7}{20} \hat{n}^{ijkl} M_{ij} M_{kl} \biggr)\biggl]\,.
\end{align}

Finally, the non-zero subleading $1/r^n$ components of the metric~\eqref{eq:gmunuNUform}, for the linear quadrupole metric and the quadratic quadrupole-quadrupole metric are given by (with ${}_{(n)}g_{ui} \equiv e_i{}^a {}_{(n)}g_{ua}$):
\begin{subequations}\label{guusub}
	\begin{align}
		\stackrel[(3)]{}{g_{uu}} &= 3 G M_{ij} \hat{n}^{ij}
 + G^2 \biggl( -  \frac{3}{5} M^{(1)}_{ij} M^{(2)}_{ij}
 -  \frac{39}{7} \hat{n}^{ij} M^{(1)}_{ik} M^{(2)}_{jk}
 -  \frac{281}{8} \hat{n}^{ijkl} M^{(1)}_{ij} M^{(2)}_{kl}\nonumber\\
& + \frac{1}{5} M_{ij} M^{(3)}_{ij}
 -  \frac{3}{7} M_{ij} \hat{n}^{ik} M^{(3)}_{kj}
 -  \frac{313}{8} M_{ij} \hat{n}^{ijkl} M^{(3)}_{kl}\biggr)\,,\\
		%%%%%%%%%%%%%%%%%%%%%%%%%%%%%%%%%%%%%%%%%%%%%%%%%%%%
		\stackrel[(4)]{}{g_{uu}} &= G^2 \biggl( \frac{4}{5} M^{(1)}_{ij} M^{(1)}_{ij}
 + \frac{12}{7} \hat{n}^{ij} M^{(1)}_{ik} M^{(1)}_{jk}
 + \frac{13}{2} \hat{n}^{ijkl} M^{(1)}_{ij} M^{(1)}_{kl}
 -  \frac{3}{5} M_{ij} M^{(2)}_{ij}\nonumber\\
& -  \frac{36}{7} M_{ij} \hat{n}^{ik} M^{(2)}_{kj}
 -  \frac{387}{8} M_{ij} \hat{n}^{ijkl} M^{(2)}_{kl}\biggr)\,,\\
		%%%%%%%%%%%%%%%%%%%%%%%%%%%%%%%%%%%%%%%%%%%%%%%%%%%%
		\stackrel[(5)]{}{g_{uu}} &= G^2 \biggl(\frac{36}{25} M_{ij} M^{(1)}_{ij}
 + \frac{54}{35} M_{ij} \hat{n}^{ik} M^{(1)}_{kj}
 -  \frac{81}{5} M_{ij} \hat{n}^{ijkl} M^{(1)}_{kl}\biggr)\,,\\
		%%%%%%%%%%%%%%%%%%%%%%%%%%%%%%%%%%%%%%%%%%%%%%%%%%%%
		\stackrel[(6)]{}{g_{uu}} &= G^2 \biggl(\frac{3}{5} M_{ij} M_{ij}
 -  \frac{21}{2} M_{ij} M_{kl} \hat{n}^{ijkl}\biggr)\,,
		%%%%%%%%%%%%%%%%%%%%%%%%%%%%%%%%%%%%%%%%%%%%%%%%%%%%
    \end{align}
\end{subequations}
and
\begin{subequations}\label{guisub}
	\begin{align}
\stackrel[(2)]{}{g_{ui}} ={}& \Bigg\lbrace 3 G M_{ij} {n}^{j}+G^2 \biggl [\biggl(- \frac{33}{20} M^{(1)}_{pq} M^{(2)}_{iq}
 -  \frac{49}{20} M^{(1)}_{iq} M^{(2)}_{pq}
 -  \frac{159}{140} M_{pq} M^{(3)}_{iq}
 -  \frac{159}{140} M_{iq} M^{(3)}_{pq}\biggr) {n}^{p}\nonumber\\
& + \biggl(- \frac{163}{8} M^{(1)}_{pq} M^{(2)}_{ij}
 -  \frac{17}{4} M^{(1)}_{pi} M^{(2)}_{qj}
 -  \frac{87}{8} M_{pq} M^{(3)}_{ij}
 -  \frac{51}{4} M_{pi} M^{(3)}_{qj}\biggr) \hat{n}^{pqj}\biggl]\Bigg\rbrace^{\mathrm{T}} ,\\
\stackrel[(3)]{}{g_{ui}} ={}&G^2 \biggl [\biggl(\frac{24}{25} M^{(1)}_{iq} M^{(1)}_{pq}
 -  \frac{53}{25} M_{pq} M^{(2)}_{iq}
 -  \frac{113}{25} M_{iq} M^{(2)}_{pq}\biggr) {n}^{p}\nonumber\\
& \qquad+ \biggl(\frac{42}{5} M^{(1)}_{pi} M^{(1)}_{qj}
 -  \frac{253}{10} M_{pq} M^{(2)}_{ij}
 -  \frac{89}{5} M_{pi} M^{(2)}_{qj}\biggr) \hat{n}^{pqj}\biggl]^{\mathrm{T}} ,\\
\stackrel[(4)]{}{g_{ui}} ={}&G^2 \biggl [\biggl(- \frac{9}{5} M_{pq} M^{(1)}_{iq}
 -  \frac{1}{5} M_{iq} M^{(1)}_{pq}\biggr) {n}^{p}
 + \biggl(- \frac{91}{8} M_{pq} M^{(1)}_{ij} + \frac{11}{8} M_{pi} M^{(1)}_{qj}\biggr) \hat{n}^{pqj}\biggl]^{\mathrm{T}} ,\\
\stackrel[(5)]{}{g_{ui}} ={}&G^2 \bigg[- \frac{42}{25} M_{iq} M_{pq} {n}^{p}
 -  \frac{51}{5} M_{pi} M_{qj} \hat{n}^{pqj}\bigg]^{\mathrm{T}} \, ,
 \end{align}
\end{subequations}
where the superscript T refers to transverse projection, \emph{i.e.}, $X^{\mathrm{T}}_i \equiv \,\perp^{ij} X_j$. For the subleading terms in the angular part of the metric, we find
\begin{subequations}\label{gabsub}
    \begin{align}
\stackrel[(2)]{}{E_{ij}} &= 2 G M_{ij}+ G^2 \biggl [ -  \frac{8}{3} M^{(1)}_{ip} M^{(2)}_{jp}
 -  \frac{52}{63} M_{ip} M^{(3)}_{jp}+\hat{n}^{pq}\biggl(\frac{1}{2} M^{(1)}_{ij} M^{(2)}_{pq}
 - 8 M^{(1)}_{pi} M^{(2)}_{qj}\nonumber\\
&\quad 
 - 3 M^{(1)}_{pq} M^{(2)}_{ij}
 -  \frac{11}{6} M_{ij} M^{(3)}_{pq}-  \frac{22}{3} M_{pi}  M^{(3)}_{qj}
 -  \frac{4}{3} M_{pq} M^{(3)}_{ij}\biggr)\biggl]^{\mathrm{TT}}\,, \label{E2}\\
\stackrel[(3)]{}{E_{ij}} &=  G^2 \biggl [- 2 M_{ip} M^{(2)}_{jp}+\hat{n}^{pq}\bigl(6 M^{(1)}_{pi} M^{(1)}_{qj}
+ M^{(1)}_{pq} M^{(1)}_{ij}
 -  \frac{9}{2} M_{ij} M^{(2)}_{pq} -  \frac{57}{2} M_{pi}  M^{(2)}_{qj}\nonumber \\ 
 &\quad-  \frac{15}{4} M_{pq} M^{(2)}_{ij}\bigr) \biggl]^{\mathrm{TT}}, \label{E3}\\
\stackrel[(4)]{}{E_{ij}} &=G^2 \biggl[\hat{n}^{pq}\biggl(\frac{173}{20} M_{ij}  M^{(1)}_{pq}
 -  \frac{67}{5} M_{pi}  M^{(1)}_{qj}
 -  \frac{97}{20} M_{pq}  M^{(1)}_{ij}\biggr) -  \frac{64}{15} M_{ip} M^{(1)}_{jp}\biggr]^{\mathrm{TT}},\\
\stackrel[(5)]{}{E_{ij}} &=   G^2 \biggl [\hat{n}^{pq}\biggl(- \frac{31}{2} M_{pi} M_{qj}
 + \frac{3}{2} M_{pq} M_{ij}\biggr) 
 -  \frac{11}{3} M_{ip} M_{jp}\biggl]^{\mathrm{TT}}\, .
    \end{align}
\end{subequations}
We recall that the TT projection is $X^{\mathrm{TT}}_{ij} \equiv \,\perp_\text{TT}^{ijmn} \!X_{mn}$, with $\perp_\text{TT}^{ijmn} = \perp^{m(i}\perp^{j)n}-\!\frac{1}{2}\perp^{ij}\perp^{mn}$.

The interaction between the mass monopole $M$ and quadrupole $M_{ij}$ leads to tail integrals, which appeared already in the waveform, through the radiative moment $U_{ij}$ in eq.~\eqref{Uijtail}, but also in the Bondi aspects $m$, $N_i$, as well as the subdominant terms ${}_{(n)}g_{uu}$ and ${}_{(n)}E_{ij}$. The complete NU metric for the interaction $M \times M_{ij}$ has been obtained in eqs.~(4.11)--(4.13) of~\cite{Blanchet:2020ngx} in ``exact'' form, valid for $r$ outside the domain of the source. The expansion of this metric at future null infinity is regular (under our assumption of past stationarity) and can be straightforwardly computed. We must split the moment into a constant piece $M_{ij}(-\mathcal{T})$, where $-\mathcal{T}$ is the finite instant before which the multipole moment is constant, and a dynamical part $M_{ij}(u)-M_{ij}(-\mathcal{T})$, which is zero in the past. The result for the integral entering the $uu$ component of the metric can be found in eq.~(4.12) of~\cite{Blanchet:2020ngx}, namely
\begin{subequations}
\begin{align}\label{eq:regular}
	&\int_{0}^{+\infty} \!\dd z \, \frac{M_{ij}(u-z)}{\left(1+\frac{z}{2r}\right)^2} =  2 r M_{ij}(-\mathcal{T}) \\& \qquad\qquad\qquad + \sum_{p=0}^{+\infty} \frac{(-)^p (p+1)}{(2r)^{p}} \int_0^{+\infty}\!\dd z \,z^{p} \Bigl[M_{ij}(u-z) - M_{ij}(-\mathcal{T})\Bigr]\,.\nn
\end{align}
Furthermore, we provide here the regular expansions of the integrals appearing in the $ua$ and $ab$ components:
\begin{align}
	& \int_{0}^{+\infty}\dd z\,\frac{5+\frac{3z}{2r}}{\left(1+\frac{z}{2r}\right)^3}M_{ij}(u-z) = 8 r M_{ij}(-\mathcal{T}) \\& \qquad\qquad\qquad + \sum_{p=0}^{+\infty} \frac{(-)^p(p+1)(p+5)}{(2 r)^p} \int_{0}^{+\infty}\dd z \,z^p \Bigl[M_{ij}(u-z)-M_{ij}(-\mathcal{T})\Bigr]\, ,\nn\\
& \int_{0}^{+\infty}\dd z\,\frac{18+\frac{8z}{r}+\frac{z^2}{r^2}}{(1+\frac{z}{2r})^4}M_{ij}(u-z) = 20 r M_{ij}(-\mathcal{T}) \\& \qquad\qquad\qquad + \sum_{p=0}^{+\infty} \frac{(-)^p(p+1)(p+3)(p+6)}{(2 r)^p} \int_{0}^{+\infty}\dd z \,z^p \Bigl[M_{ij}(u-z)-M_{ij}(-\mathcal{T})\Bigr]\, .\nn
\end{align} 
\end{subequations}
These expansions confirm that under the assumption of stationarity in the past, the radial expansion is regular to any order. In particular, the latter integral will contribute to the subdominant ${}_{(n)}{E_{ij}}$ for any $n \geqslant 3$. Explicitly, the $M \times M_{ij}$ contributions to $m$, $N_i$, $H^{TT}_{ij}$ and ${}_{(n)}{E_{ij}}$ are given by
\begin{subequations}\label{Entail}
\begin{align}
m\big\vert_{M \times M_{ij}}&=6G^2 M n^{ij} \int^{+\infty}_0 dz \left[ \ln\left(\frac{z}{2b_0}\right)+\frac{11}{12}\right] M_{ij}^{(4)}(u-z)\,, \\
N_i\big\vert_{M \times  M_{ij}}&=12 G^2 M  \perp_{ik}n^j \int^{+\infty}_0 dz \left[ \ln\left(\frac{z}{2b_0}\right)+\frac{11}{12}\right] M_{kj}^{(3)}(u-z)\,, \\ 
H^{\mathrm{TT}}_{ij}\big\vert_{M \times  M_{ij}}&= 4 G^2 M \perp^{\mathrm{TT}}_{ijkl}\int^{+\infty}_0 dz \left[ \ln\left(\frac{z}{2b_0}\right)+\frac{11}{12}\right] M_{kl}^{(4)}(u-z)\,,\\
\stackrel[(2)]{}{E_{ij}}\Big\vert_{M \times M_{ij}}&= G^2 M \perp^{\mathrm{TT}}_{ijkl}\left( M^{(1)}_{kl} + 4 \int_0^{+\infty} \left[ \ln\left(\frac{z}{2b_0}\right)+\frac{11}{12}\right] M_{kl}^{(2)}(u-z)\right)\,, \\
\stackrel[(3)]{}{E_{ij}}\Big\vert_{M \times M_{ij}}&= 5G^2M \perp^{\mathrm{TT}}_{ijkl} M_{kl}(-\mathcal T)\,, \label{Entaile}\\
\stackrel[(n \geqslant 4)]{}{E_{ij}}\Big\vert_{M \times  M_{ij}}&= G^2 M   \perp^{\mathrm{TT}}_{ijkl}\frac{(-)^n(n-3)(n-1)(n+2)}{2^{n-2}} \int_{0}^{+\infty}\dd z \,z^{n-4} \Bigl[M_{kl}(u-z)-M_{kl}(-\mathcal{T})\Bigr]\,.
\end{align} 
\end{subequations}

\section{Bondi aspects and charges}
\label{sec:IR}

In this section, we will discuss gravitational charges in the multipolar expansion of the metric. We will discuss two classes of gravitational charges: 1) those strictly conserved in the generic dynamics of GR in asymptotically flat spacetimes, 2) charges that are time-independent only in non-radiative spacetimes, while they obey a flux-balance equation in the presence of radiation. For the first class, we will discuss the Newman-Penrose charges \cite{Newman:1965ik,Newman:1968uj}, while for the latter, we will study the celestial charges as defined in \cite{Grant:2021hga,Freidel:2021ytz,Compere:2022zdz}. 

\subsection{Newman-Penrose charges}

The Newman-Penrose (NP) charges \cite{Newman:1965ik,10.2307/2415610,Newman:1968uj,Exton:1969im} are exactly conserved quantities at any $u$ defined at null infinity. In the MPM formalism, assuming that there is no incoming radiation from  past null infinity, the metric is entirely determined as a functional of the canonical multipole moments $M_L(u)$, $S_L(u)$, which can have arbitrary time dependence, except for the lowest $\ell=0,1$ multipole moments: $M,\, M_i \equiv G_i = P_i \, u + K_i$ and $S_i$.  Under the stronger assumption of past stationarity, all multipole moments become constants before a given retarded time $u=-\mathcal T$. Thus, $P_i=0$ and, furthermore, in the center-of-mass frame that we are using, $M_i=0$. Therefore, a fully conserved quantity is expected to be a polynomial of the set of Geroch-Hansen multipole moments at spatial infinity given in terms of $M_L(-\mathcal T)$, $\ell \geqslant 0$, $S_L(-\mathcal T)$, $\ell \geqslant 1$. In what follows, we evaluate the NP charges and check that they do not display quadrupole-quadrupole interactions, while they reproduce known expressions including monopole-quadrupole tail interactions.

The NP charges are defined using the Weyl scalar 
\begin{align}
    \Psi_0=-C_{\mu \nu \alpha \beta}\, \ell^\mu m^\nu \ell^\alpha m^\beta\,,
\end{align}
in terms of a null tetrad, consisting of the incoming and outgoing real null vectors
\begin{align}
    n=e^{-2 \beta}
    \bigl(\partial_u-F \partial_r+U^a \partial_a\bigr)\,, \quad \ell=\partial_r\,, 
\end{align}
and the complex null vector $m$ and its complex conjugate $\bar m$, with
\begin{align}
    m=\frac{1}{r}\left( \zeta^a-\frac{1}{2r}C^{a}{}_{b}\zeta^b+\mathcal{O}(r^{-2})\right)\partial_a +\omega \partial_r\,.
\end{align}
In the above equations, $\beta$, $F,U^a$, $\zeta^a$ and $\omega$ are functionals of the Bondi data that are fixed by the normalization property of the tetrad (see \emph{e.g.},~\cite{Barnich:2019vzx,Freidel:2021fxf,Seraj:2021rxd,Seraj:2022qyt}).  The pair $(\zeta^a,\bar\zeta^a)$ forms a null dyad on the sphere normalized as $\gamma_{ab}\zeta^a\bar\zeta^b=1$. The bulk extension of $m$ is fixed by requiring that it is parallelly transported along the outgoing null vector, \emph{i.e.} $\ell^\nu\nabla_\nu m^\mu=0$, as in \cite{Newman:1968uj}. One may choose the dyad to be adapted to the stereographic coordinates $(z,\bar z)$ on the sphere, in which the metric reads $\dd s^2=\frac{4}{(1+z\bar z)^2}\dd z \dd\bar z$, so that $\zeta^a \pd_a=\frac{1+z\bar z}{2}\pd_z$.  
Residual transformations of the tetrad only include global Type III rotations that covariantly transform $\zeta^a$.\footnote{This residual symmetry was used in \cite{Godazgar:2022pbx} to interpret the gyroscopic memory as a vacuum transition.}  Computing the asymptotic form of this Weyl scalar, we find that $\Psi_0={\Psi_0^0}\,r^{-5}+{\Psi_0^1}\,r^{-6}+\cO(r^{-7})$, where the leading and subleading coefficients are 
\begin{align}
    \Psi_0^0&=3\Big(\!\stackrel[(2)]{}{{E}_{ab}}-\frac{1}{16}C^2 C_{ab}\Big)\zeta^a\zeta^b\,,\qquad \Psi_0^1=6\stackrel[(3)]{}{{E}_{ab}} \zeta^a\zeta^b\,. 
\end{align}
The (complex-valued) Newman-Penrose charges are defined, for $m=-2,-1,0,1,2$, as
\begin{align}
    Q_m &\equiv \oint_S  {}_2\overline{Y}_{2m}\Psi_0^1\,,
\end{align}
where ${}_s{Y}_{lm}$ denote the spin weighted spherical harmonics of spin weight $s$ (the bar indicates the complex conjugate). They are given for $0 \leqslant s \leqslant l$ in terms of the Geroch-Held-Penrose operator $\eth$\footnote{Recall that the $\eth$ and $\bar\eth$ operators act on a weighted scalar $W=W_{a_1\cdots a_n b_1\cdots b_m}\zeta^{a_1}\cdots \zeta^{a_n}\bar\zeta^{b_1}\cdots \bar\zeta^{b_m}$ as 
\begin{align*}
\eth W=\zeta^c \zeta^{a_1}\cdots \zeta^{a_n}\bar\zeta^{b_1}\cdots \bar\zeta^{b_m} D_c W_{a_1\cdots a_n b_1\cdots b_m},\\
\bar\eth W=\bar \zeta^c \zeta^{a_1}\cdots \zeta^{a_n}\bar\zeta^{b_1}\cdots \bar\zeta^{b_m} D_c W_{a_1\cdots a_n b_1\cdots b_m}.
\end{align*} 
} 
by 
\begin{subequations}
\begin{align}
	{}_s{Y}_{lm} &= \sqrt{\frac{(l-s) !}{(l+s) !}} \,\eth^s Y_{l m}\,,\\
	\text{and thus}\quad {}_2\overline{Y}_{2m} &= \frac{1}{2\sqrt{6}}\,\bar{\eth}^2 \overline{Y}_{2 m}=\frac{1}{2\sqrt{6}}\,\bar{\zeta}^a\bar{\zeta}^b D_a D_b \overline{Y}_{2 m}\,.
\end{align}
\end{subequations}
Upon using the decomposition $\zeta^a \bar{\zeta}_b=\frac{1}{2} \delta^a{}_b+\frac{\ii}{2} \eps^a{}_b$ and integrating by parts, we find 
\begin{align}
    Q_m&=\frac{1}{2}\sqrt{\frac{3}{2}}\oint_S  \overline{Y}_{2 m}D_aD_b\,\Big(\!\stackrel[(3)]{}{{E}^{ab}} - \ii\stackrel[(3)]{}{\widetilde{E}^{ab}}\Big)\,,
\end{align}
where, for any STF tensor $X_{ab}$, we pose $\widetilde{X}_{ab}=\eps_{ac}X^c{}_b$. Now, we make a change of basis from spherical harmonics $Y_{2m}$ to STF harmonics $\hat{n}_{ij}$, and we adjust the normalization to define STF NP charges $Q_{ij}$ as
\begin{align}\label{NP charge simplified}
    Q_{ij}&\equiv \oint_S D^aD^b\hat{n}_{ij}\,\Big(\!\stackrel[(3)]{}{{E}_{ab}}-\ii\stackrel[(3)]{}{\widetilde{E}_{ab}}\Big)\,.
\end{align}
Given that $D_a n^i=e_a{}^i$ and $D_aD_b n^i=-\gamma_{ab}n^i$, it can be checked that $D^aD^b\hat{n}_{ij}=2e^a{}_{\langle i} e^b{}_{j\rangle}$. Therefore, we arrive at
\begin{align}
   Q_{ij}& =2\oint_S  \;e^a{}_{\langle i} e^b{}_{j\rangle}\Big(\!\stackrel[(3)]{}{{E}_{ab}}-\ii\stackrel[(3)]{}{\widetilde{E}_{ab}}\Big) = 2\oint_S \Big(\!\stackrel[(3)]{}{{E}_{ij}}-\ii\stackrel[(3)]{}{\widetilde{E}_{ij}}\Big)\,, \label{eq: NP charge}
\end{align}
where we have resorted to the Cartesian presentation of the Bondi data 
\begin{align}
\stackrel[(3)]{}{{E}_{ij}}\equiv e^a{}_{ i} \,e^b{}_{j}\stackrel[(3)]{}{{E}_{ab}}\,,\qquad\stackrel[(3)]{}{{\widetilde E}_{ij}}\equiv e^a{}_{ i} \, e^b{}_{j}\stackrel[(3)]{}{{\widetilde E}_{ab}}=n_m\eps_{min}\stackrel[(3)]{}{E_{nj}} \,.
\end{align}
Evaluating the integral in eq.~\eqref{eq: NP charge} amounts to computing the monopole moment of its integrand. However, by explicit computation of the TT projection in eq.~\eqref{E3}, we observe that neither ${}_{(3)}{E}_{ij}$ nor ${}_{(3)}\widetilde{E}_{ij}$ contain monopole terms. Thus, there is no mass quadrupole-quadrupole contribution to the NP charges. The mass monopole-quadrupole contribution to the NP charges is a direct consequence of eqs.~\eqref{Entail} obtained from eq.~(4.13b) of~\cite{Blanchet:2020ngx}.
Expanding $\perp^{\mathrm{TT}}_{ijkl}$ in the expression of ${}_{(3)}{E_{ij}}$ in eq.~\eqref{Entaile} one finds 
\begin{align}
	Q_{ij} &= 5G^2 M \oint_S \Bigl[2M_{ij}(-\mathcal{T})-2M_{il}(-\mathcal{T})n_jn_l-2M_{jl}(-\mathcal{T})n_{il} + n_{ikjl} M_{kl}(-\mathcal{T}) \Bigr] \nn\\
	&= 4 G^2 MM_{ij}(-\mathcal{T})\,.\label{NPcharge}
\end{align}
The NP charge is indeed conserved for any $u$.
This result is consistent with the expressions found by Newman-Penrose \cite{Newman:1965ik,Newman:1968uj} and van den Burg \cite{10.2307/2415610}. Note that our result is written in the center-of-mass frame in which the mass dipole moment is zero.

\subsection{From Newman-Unti to Bondi coordinates}
\label{sec:NUBondi}

In Section~\ref{sec:quadratic}, we obtained the radiative metric corresponding to selected multipole interactions in Newman-Unti (NU) gauge. However, many results on the asymptotic structure of asymptotically flat spacetimes are known in Bondi gauge instead. In the following, we complete the map from NU to Bondi gauge outlined in App.~A of~\cite{Blanchet:2020ngx}. 

The NU and Bondi coordinates are related by a change of radius, namely  $r_\text{NU}$ in NU coordinates and $r_\text{B}$ in Bondi coordinates, with $u$ and angle coordinates $\theta^a$ unchanged. In Bondi gauge, the spherical metric takes the form~\cite{Grant:2021hga}
\begin{equation}
	g_{ab}=r_\text{B}^2 \sqrt{1+\frac{\mathcal C^{\text{B}}_{cd}\mathcal C_\text{B}^{cd}}{2r_\text{B}^2}}\gamma_{ab}+r_\text{B}\,\mathcal C^{\text{B}}_{ab}\,,\qquad  \mathcal C^{\text{B}}_{ab} \equiv e^{i}{}_{\langle a} e^{j}{}_{b\rangle}\left(H^{\mathrm{TT}}_{i j}+\sum_{n=2}^{+\infty}\frac{1}{r_\text{B}^{n}}\stackrel[(n)]{}{E^\text{B}_{ij}}\right)\,. 
\end{equation} 
The relation between the two radii follows from 
\begin{equation}\label{rBNU}
	r_\text{B}^4 =\left. \frac{\text{det}\; g_{ab}}{\text{det}\; \gamma_{ab}}\right\vert_\text{NU} \,.
\end{equation}
We shall parametrize the NU metric as
\begin{equation}
	 g_{ab}=\Bigl(r_\text{NU}^2+W(r_\text{NU},\theta^a)\Bigr) \gamma_{ab}+ r_\text{NU} \,\mathcal C^\text{NU}_{\langle a b \rangle}(r_\text{NU},\theta^a)\,,
\end{equation}
where $W=\mathcal{O}(G^2)$ and $\mathcal{C}^\text{NU}_{\langle a b \rangle}=\mathcal{O}(G)$. Then, eq.~\eqref{rBNU} yields 
\begin{align}
	r_\text{B} &= \Bigl[\bigl( r_\text{NU}^2 + W\bigr)^2 - \frac{ r_\text{NU}^2}{2}\mathcal C^\text{NU}_{\langle a b \rangle} \mathcal C_\text{NU}^{\langle a b \rangle} \Bigr]^{1/4}= r_\text{NU} +\frac{1}{2r_\text{NU}}  \left(W-\frac{1}{4}\mathcal C^\text{NU}_{\langle a b \rangle} \mathcal C_\text{NU}^{\langle a b \rangle}\right)+\mathcal{O}(G^4)\,, \label{exprB}
\end{align}
where $W$ and $\mathcal C^\text{NU}_{\langle a b \rangle} \mathcal C_\text{NU}^{\langle a b \rangle}$ can be evaluated either at radius $r_\text{NU}$ or $r_\text{B}$, neglecting $\mathcal{O}(G^4)$ corrections. At large radius, it can be shown, using $g_{ur}^\text{B}(u,r_B,\theta^a) = -\partial r_\text{NU}/\partial r_\text{B}$ and $g_{ab}^\text{B}(u,r_\text{B},\theta^a)=g_{ab}^\text{NU}(u,r_\text{NU},\theta^a)$, combined with the $1/r_\text{B}$ expansion of $g_{ur}^\text{B}$, that 
\begin{equation}
	W=\frac{1}{8}C_{ab}C^{ab}+\mathcal{O}(r_\text{NU}^{-1})\,,\qquad  \mathcal C^\text{NU}_{\langle ab \rangle }=C_{ab}+ \mathcal{O}(r_\text{NU}^{-1})\,, 
\end{equation} 
and therefore 
\begin{equation}
	r_\text{B}=r_\text{NU}-\frac{1}{16 r_\text{NU}} C_{ab}C^{ab}+\mathcal{O}(r_\text{NU}^{-2})\,,
\end{equation} 
which reproduces eq.~(61b) of \cite{Blanchet:2020ngx}. Including all orders in the radius, we may pose 
\begin{equation}
	\mathcal C^\text{NU}_{\langle ab \rangle }=C_{ab}+e_{\langle a}^i e_{b\rangle}^j \sum_{n=2}^{+\infty} r_\text{NU}^{-n}\stackrel[(n)]{}{E_{ij}}\,.
\end{equation} 
We see that the relation~\eqref{exprB} can be computed exactly at quadratic order, knowing $W$ from the quadratic NU metric, as well as the linear part of $C_{ab}$ and ${}_{(k)}{E_{ij}}$\footnote{Note that in~\cite{Blanchet:2020ngx}, the linear part of ${}_{(k)}{E_{ij}}$ is denoted by $Q^{ij}_{k}$ for $k\geqslant 2$.}
given, respectively, by eqs.~(3.18) and (3.15c) in~\cite{Blanchet:2020ngx}. Keeping only linear terms in $M$ and $M_{ij}$ together with BMS diffeomorphisms, we first get
\begin{equation}
	\mathcal C^\text{NU}_{\langle ab \rangle} = -2 G D_{\langle a} D_{b \rangle} f + 2 G e_{\langle a}^i e_{b \rangle}^j \left(M_{ij}^{(2)}+\frac{1}{r_\text{NU}^2}M_{ij}\right)\,, 
\end{equation}
where $f=T(\theta^a)+\frac{u}{2}D_a Y^a$ results from a BMS transformation. In the absence of BMS transformation, we then find 
\begin{equation}\label{r2}
	r_\text{NU}=r_\text{B} - \frac{1}{2r_\text{B}}\biggl[W-G^2\perp_\text{TT}^{ijkl}\left(M_{ij}^{(2)}+\frac{1}{r_\text{B}^2}M_{ij}\right)\left(M_{kl}^{(2)}+\frac{1}{r_\text{B}^2}M_{kl}\right) \biggr]+\mathcal{O}(G^4)\,.     
\end{equation}
The Bondi mass aspect can be read in Bondi gauge from $g_{uu}=-1+2m/r+O(r^{-2})$. The Bondi angular momentum aspect $N_a$ can be read in Bondi gauge from\footnote{Given the non-universal conventions in the literature, we provide here a dictionary to some other references: $N_a$ as defined in \cite{Flanagan:2015pxa,Grant:2021hga} is equal to $N_a$ here; the covariant momentum $\mathcal{P}_a$ as defined in \cite{Freidel:2021fxf,Freidel:2021qpz} is equal to $N_a$ here, and $N_a$ as defined in \cite{Barnich:2010eb} is equal to $N_a-\frac{1}{4}C_{ab}D_c C^{cb}-\frac{3}{32}\p_a (C_{cd}C^{cd})$ here.} 
\begin{equation}
	g_{ua}=\frac{ D^b C_{ab}}{2}+\frac{2}{3r_\text{B}} \left[N_a-\frac{3}{32} D_a \bigl(C_{bc}C^{bc}\bigr)\right]+ \mathcal{O}(r_\text{B}^{-2})\,.
\end{equation}
Since the Bondi radius $r_\text{B}$ differs from the Newman-Unti radius $r_\text{NU}$ by terms of order $G^2$ while $H_{ij}^\text{TT}$ and $\stackrel[(n)]{}{E_{ij}}$ are of order $G$, we simply have
\begin{equation}
	\stackrel[(n)]{}{E^\text{B}_{ij}} \,=\, \stackrel[(n)]{}{E_{ij}} +\, \mathcal{O}(G^3)\,.     
\end{equation}
This completes the map of all initial data in Bondi gauge in terms of Newman-Unti initial data at order $G^2$ for the interactions considered in this work. 

\subsection{Flux-balance laws of the  dressed \texorpdfstring{$n \leqslant 3$}{} Bondi aspects}

In the Bondi expansion, Einstein's equations lead to time evolution equations for all $n \geqslant 0$ Bondi aspects. The Bondi aspects can be further ``dressed'' by adding suitable non-linear combinations of Bondi data such that the dressed Bondi aspects are identically conserved when the news vanishes. In this section, we recall the definitions of the dressed $n=0,1,2$ Bondi aspects and propose a new definition of the dressed $n=3$ Bondi aspect following the construction of \cite{Grant:2021hga}. We do not consider the case where $n >3$.

It will be useful to first introduce the gravitational electric-magnetic (or mass-current) duality covariant mass \cite{Godazgar:2018qpq,Godazgar:2018dvh,Godazgar:2019dkh}
\begin{align}
    m_{ab} &= m \,\gamma_{ab} + \frac{1}{2}D_{[a}D^c C_{b]c}\,.\label{mab}
\end{align}
Mass multipole moments are annihilated by the differential operator entering the second term of eq.~\eqref{mab}. As a result, only current multipole moments appear in eq.~\eqref{mab}. The flux-balance laws of the first $n=0,1,2,3$ Bondi aspects are given by  
\begin{subequations}\label{FBE}
\begin{align}
\partial_u m &= -\frac{1}{8}N^2 +\frac{1}{4}D_aD_bN^{ab}\,, \\
\partial_u N_a &= D^b m_{ab} +\frac{1}{4} \left(N^{bc}D_c C_{ab} + 3 C_{ab}D_cN^{bc}\right)\,, \\ \label{eq:balanceE2}
    \partial_u \stackrel[(2)]{}{E_{ab}} &= \frac{1}{4}(C N) C_{ab} + \frac{1}{2}m_{ac}C^{c}_{\;b} +\frac{1}{3}D_{\langle a}N_{b\rangle}\,,\\ \label{eq:balanceE3}
    \partial_u \stackrel[(3)]{}{{E}_{ab}} &= \mathcal{D}_0\stackrel[(2)]{}{{E}_{ab}} + D^c \left[ \left(\frac{1}{4} D_e C^{de} C_{d \langle a} - \frac{3}{32}D_{\langle a }C^2 + \frac{5}{32}C^2 D_{\langle a} - \frac{1}{3}N_{\langle a}\right)C_{ b\rangle c} \right]\,,
\end{align}
\end{subequations}
where we use the compact notation $C^2 \equiv C_{ab}C^{ab}$, $(C N) \equiv C_{ab}N^{ab}$, $N^2 \equiv N_{ab}N^{ab}$, and where $\mathcal D_0 \equiv -\frac{1}{4}(\Delta+2)$. The $n=1$ flux-balance law matches~\cite{Flanagan:2015pxa}. The evolution of the $n=2$ Bondi aspect in vacuum was first derived in \cite{10.2307/2415610} and agrees with that obtained in~\cite{Nichols:2018qac}.  The $n=2,3$ expressions are taken from~\cite{Grant:2021hga} (see alternative derivations in \cite{10.2307/2415610,Barnich:2011ty,Godazgar:2018vmm}). 

The $n=0$ Bondi aspect $m$ is just the Bondi mass aspect that permits defining the supermomenta 
\begin{equation}
    \mathcal P_L = \oint_S m \, \hat{n}_L\,, \label{PT}
\end{equation} 
\emph{i.e.}, the canonical charges associated with supertranslations $T=T_{L}\hat{n}_L$. Its explicit expression, in the post-Minkowskian approximation, restricted to linear terms, tails and quadrupole-quadrupole interactions can be obtained from eq.~\eqref{mexpl}. For the monopole $\ell=0$, 
%$L=\emptyset$, 
the energy is defined as $G \, \mathcal P_0 (u) \equiv \mathcal E(u)=M - E^{\text{GW}}(u)$, 
where $E^{\text{GW}}$ was defined in eq.~\eqref{fluxE}.

The $n=1$ Bondi aspect $N_a$, usually referred to as the angular momentum aspect, can be supplemented or ``dressed'' with suitable terms such that its retarded time derivative vanishes when the news vanishes. The dressed $n = 1$~\cite{Compere:2018ylh,Compere:2019gft} and $n=2$ Bondi aspects~\cite{Grant:2021hga} read as\footnote{Equation (2.10) of~\cite{Compere:2022zdz} should read as eq.~\eqref{curlyE2}.}
\begin{subequations}\label{curlyNaE2}
\begin{align}
    \mathcal{N}_a &= N_a-\frac{1}{16} D_a C^2-\frac{1}{4}C_{a}^{\;b}D^c C_{bc} - u D^b m_{ab}\,, \\ 
\stackrel[(2)]{}{\mathcal{E}_{ab}}  &=\,\stackrel[(2)]{}{E_{ab}}-\frac{u}{2}C^c_{(a}m_{b)c}-\frac{u}{3}D_{\langle a} N_{b \rangle}+\frac{u^2}{6}D_{\langle a} D^c m_{b \rangle c}\,.\label{curlyE2}
\end{align}
\end{subequations}
They obey the flux-balance laws 
\begin{subequations}\label{fb}
\begin{align}
\partial_u \mathcal N_a & =  \frac{1}{4}N^{bc}D_bC_{ca}+\frac{1}{2}C_{ab}D_c N^{bc}- \frac{1}{4}N_{ab}D_c C^{bc} -\frac{1}{8}D_a (CN)+\frac{u}{8} D_a N^2  -\frac{u}{2}D_c D_{\langle a}D_{b\rangle} {N}^{bc}\,, \label{FBL1}\\
 \partial_u  \stackrel[(2)]{}{\mathcal E_{ab}} &=\frac{1}{4}(CN)C_{ab}- \frac{u}{2}\partial_u (C_{a}^{\; c}m_{bc})-uD_{\langle a}\left(\frac{1}{12} N^{cd}D_{\underline{d}}C_{\underline{c} b \rangle}+ \frac{1}{4}C_{b\rangle c}D_d N^{cd}\right) -\frac{u^2}{48}D_{\langle a} D_{b\rangle}N^2\nonumber\\
 &\quad + \frac{u^2}{12}\text{STF}_{ab}\Bigl[D_a D_c D_{\langle b}D_{d\rangle}  {N}^{cd}\Bigr]\,. \label{FBL2}
\end{align}
\end{subequations}
The $n=1$ dressed aspect allows defining the super-Lorentz conserved charges 
\begin{equation}
\mathcal J_L = -\frac{1}{2}\oint_S \eps^{ab}\partial_b \hat n_L \mathcal N_a, \qquad \mathcal K_L = \frac{1}{2}\oint_S \gamma^{ab}\partial_b \hat n_L \mathcal N_a \label{JL}
\end{equation}
for arbitrary $\text{Diff}(S^2)$ generators $Y^a=-\eps^{ab}\partial_b \hat n_L+\gamma^{ab}\partial_b \hat n_L$. The $n=2$ dressed aspect is associated with the $n=2$ ``celestial'' charges $\oint_S {}_{(2)}\mathcal{E}_{ab} (D^a D^b \mathcal{S}^+ +\eps_{ac}D^b D^c \mathcal{S}^-)$, defined for arbitrary scalars $\mathcal{S}^\pm(\theta^a)$ on the 2-sphere.

The dressed $n=3$ Bondi aspect, which is conserved in non-radiative regions, was introduced in eq.~(4.43) of~\cite{Grant:2021hga}.\footnote{However, three typos arose in writing this expression as can be deduced from logically following the algorithm described in \cite{Grant:2021hga}; see erratum \cite{PhysRevD.107.109902}. In eq.~(4.40c), the factor $1/12$ should read $1/6$; in eq.~(4.39a), the factor $-3/4$ should be $-3/8$ and the factor $5/4$ should be $5/8$.} It can be written in terms of the  duality-covariant quantity $m_{ab}$ as it should come from gravitational electric-magnetic duality \cite{Compere:2022zdz}. The (corrected)  dressed $n=3$ Bondi aspect reads as 
\begin{align}\label{curvyE3}
   \stackrel[(3)]{}{\mathcal{E}_{ab}} \,=\, \stackrel[(3)]{}{E_{ab}} &- u \left\{\mathcal D_0\stackrel[(2)]{}{E_{ab}} + D^c \left[ \left( \frac{1}{4} D_e C^{de} C_{d \langle a}-\frac{3}{32}D_{\langle a }C^2 +\frac{5}{32}C^2 D_{\langle a} -\frac{1}{3}N_{\langle a}\right)C_{ b\rangle c} \right]\right\} \nonumber\\
   & \hspace{-1cm} +\frac{u^2}{2} \left[-\frac{1}{3}D^c\left(D^d m_{d \langle a}C_{b\rangle c}\right)+\frac{1}{2}\mathcal D_0\left(m_{ac} C^c_{\; b}\right)+\frac{1}{3}\mathcal D_0 D_{\langle a} N_{b\rangle} \right] - \frac{u^3}{18}\mathcal D_0 D_{\langle a}D^cm_{b\rangle c}\,.
\end{align}
It obeys the flux-balance law
\begin{equation}\label{FBL3}
  \partial_u\stackrel[(3)]{}{\mathcal{E}_{ab}} \,=\, \stackrel[(3)]{}{\mathcal{F}_{ab}}   -\frac{u^3}{36}\mathcal{D}_0\,\text{STF}_{ab}\left(D_a D_c D_{\langle b}D_{d\rangle}N^{cd}\right) ,
\end{equation}
where the flux ${}_{(3)}{\mathcal{F}_{ab}}$ can be written, using the flux-balance laws \eqref{FBE} and the identity $C_{c(a}N^c_{\;\; b)}=\frac{1}{2}\gamma_{ab}(CN)$, as
\begin{align}\label{F3dressed}
    \stackrel[(3)]{}{\mathcal{F}_{ab}} \,=&  -\frac{u}{4}\mathcal{D}_0 \left((CN) C_{ab}\right) + \frac{u}{3}D^c \left[ N_{\langle a}N_{b\rangle c} +\left(\frac{1}{4}N^{de}D_e C_{d\langle a}+\frac{3}{4}D_e N^{de} C_{d\langle a}\right)C_{b\rangle c}\right] \nonumber\\
    &+\frac{u}{4}\partial_u D^c \left[\frac{3}{8} D_{\langle a}\left(C^2 C_{b\rangle c}\right) -   \left(D_e C^{de}\right)\left(C_{d\langle a}C_{b\rangle c}\right) - C^2D_{\langle a} C_{b\rangle c}\right]\nonumber \\
    &+\frac{u^2}{24}\mathcal{D}_0 D_{\langle a}\left(N^{de}D_{\underline e} C_{b\rangle d} + 3 C_{b\rangle d}D_e N^{de}\right)+\frac{u^2}{2}\partial_u\left[-\frac{1}{3}  D^c \left(D^d m_{d \langle a}C_{b\rangle c}\right) +  \frac{1}{2} \mathcal{D}_0 \left(m_{a c}C^{c}_{\;b}\right) \right] \nonumber\\
    & +\frac{u^3}{144}\mathcal{D}_0 D_{\langle a}D_{b\rangle} N^2\,.
\end{align}
As expected, it vanishes when the news does. We cross-checked with a computer code that the flux-balance laws for ${}_{(2)}\mathcal{E}_{ab}$ and ${}_{(3)}\mathcal{E}_{ab}$ are indeed obeyed for the $M_{ij}\times M_{ij}$ interactions.

\paragraph{Use of dimensional identities.} In order to verify the flux-balance laws in practice (and, more generally, to verify the relations between different pieces of the radial expansion of the metric), we must make use of so-called dimensional identities. Indeed, in some cases, the difference between the left and right-hand sides of the balance equations, although zero, is not \textit{manifestly} zero. Indeed, given a tensor of rank greater than four in dimension three, say $T_{ijkl}$, an identity such as $T_{[ijkl]} \equiv 0$, referred to as dimensional identities, is not explicitly ``apparent''.

In our problem, we meet expressions like $M^{(n_1)}_{ij}M^{(n_2)}_{kl} \hat{n}_{mL} e_a^p e_b^q$ which do contain at least four free indices whose antisymmetrization will not trivially yield zero, \emph{e.g.}, the indices $i$, $k$, $m$ and $p$ in this example. Moreover, the number of free indices can be reduced by contracting pairs lying outside the antisymmetrization operator $[\cdots]$. In particular, it is possible to construct a rank 2 dimensional identity from the latter monomial, \emph{e.g.}, by contracting the pairs $\{i,j\}$, $\{l,m\}$, $\{p,q\}$, $\{k,L=n\}$ (assuming that the length of $L$ is 1). The dimensional identities produced in that way are thus
\begin{align}\label{dim identities}
	\stackrel{(n_1,n_2)}{I_{ab}} \equiv -2 G^2 M^{(n_1)}_{j[i}M^{(n_2)}_{k\underline{l}} \hat{n}_{m\underline{n}} e_a^{p]}\, e_b^q \, \delta^{ij}\, \delta^{lm}\, \delta_{pq}\, \delta^{kn} = 0\, .
\end{align}
They are used in several instances of our calculations.  For example, we find that the TT projection of the piece of $g_{ab}$ that is proportional to $r^0$ is
\begin{align}
	8 \!\stackrel{(1,3)}{I_{ab}} + 2 \!\stackrel{(0,4)}{I_{ab}} = 0\, .
\end{align}
The differences of both sides of the balance equations~\eqref{eq:balanceE2} for ${}_{(2)}E_{ab}$ and~\eqref{eq:balanceE3} for ${}_{(3)}E_{ab}$ reduce, respectively, to the identities
\begin{subequations}
	\begin{align}
		& 8 \!\stackrel{(1,3)}{I_{ab}} + 2 \!\stackrel{(0,4)}{I_{ab}} - 6 \!\stackrel{(2,2)}{I_{ab}} =0\, , \\ 
		& 21 \!\stackrel{(1,2)}{I_{ab}} + 23 \!\stackrel{(0,3)}{I_{ab}} =0 \, .
	\end{align}
\end{subequations}
Finally, subtracting the right-hand sides from the left-hand sides of the balance equations~\eqref{FBL2} and~\eqref{FBL3} for the dressed aspects  ${}_{(2)}\mathcal{E}_{ab}$ and ${}_{(3)}\mathcal{E}_{ab}$, respectively, lead to
\begin{subequations}
	\begin{align}
		& 8  \!\stackrel{(1,3)}{I_{ab}} + 2 \!\stackrel{(0,4)}{I_{ab}} - 6 \!\stackrel{(2,2)}{I_{ab}}=0 \, , \\
		& 21 \!\stackrel{(1,2)}{I_{ab}} + 23  \!\stackrel{(0,3)}{I_{ab}} - 4 u \!\stackrel{(1,3)}{I_{ab}} - u \!\stackrel{(0,4)}{I_{ab}} + 3 u \!\stackrel{(2,2)}{I_{ab}} = 0\, .
	\end{align}
\end{subequations}
Hence thanks to the identities \eqref{dim identities}, we have been able to verify the above claims. A way to by-pass the use of dimensional identities is to specify and expand all the components of vectors and tensors in a given chosen basis. 

\subsection{The \texorpdfstring{$n\leqslant 3$}{} celestial charges at \texorpdfstring{$G^2$}{} order}

In the linear theory, the dressed $n \geqslant 0$ Bondi aspects have been shown to provide the complete set of conserved charges for asymptotically flat spacetimes admitting a Bondi expansion~\cite{Compere:2022zdz}. In this section, we will explicitly write the dressed $n \leqslant 3$ Bondi aspects in the post-Minkowskian approximation, with terms linear in $M$ and $M_{ij}$ and $M \times M_{ij}$ and $M_{ij}\times M_{kl}$ interactions, and compute the corresponding charges.

For convenience, we shall express the Bondi data in terms of the corresponding Cartesian transverse tensors
\begin{align}\label{TT Bondi data}
    \mathcal{N}_{i}\equiv e^{a}{}_{i}\,\mathcal{N}_{a}\,, \qquad \stackrel[(n)]{}{\mathcal{E}_{ij}}=e^{a}{}_{ i} e^{b}{}_{j}\stackrel[(n)]{}{\mathcal{E}_{ab}}\,.
\end{align}
Summing up the different contributions in eqs.~\eqref{curlyNaE2} and \eqref{curvyE3} [with the undressed quantities therein given by eqs.~\eqref{Naexpl}, \eqref{E2}, and \eqref{E3}, respectively], we obtain through computer calculation the explicit expressions for $\mathcal{N}_{i}{}$, ${}_{(2)}\mathcal{E}_{ij}$ and ${}_{(3)}\mathcal{E}_{ij}$. Note that ${}_{(n)}\mathcal{E}_{ij}$ is traceless with respect to $\perp_{ij}$ because ${}_{(n)}\mathcal{E}_{ab}$ is traceless with respect to $\gamma_{ab}$. Given the length of the resulting expressions for~\eqref{TT Bondi data}, we only provide their expressions for quadrupole-quadrupole interactions in Appendix~\ref{app:B}. 

We are now in the position to compute the contributions up to order $G^2$ to the Bondi supermomenta, super-angular momenta and super-center-of-mass defined in eqs.~\eqref{PT} and \eqref{JL} as well as the two celestial charges $\mathcal Q^{\pm}_{2,L}$ and $\mathcal Q^{\pm}_{3,L}$ defined, respectively, from the $n=2$ and $n=3$ Bondi aspects. The latter charges, which are STF with respect to their indices $L$, are defined in~\cite{Compere:2022zdz} by
\begin{equation}\label{celestial charges def}
    \mathcal Q^+_{n,L} \equiv \oint_{S} \stackrel[(n)]{}{\mathcal{E}^{ab}} D_{a}D_b \hat n_L \,,\qquad \mathcal Q^-_{n,L} \equiv \oint_{S} \stackrel[(n)]{}{\mathcal{E}^{ab}} \eps_{ac} D_{b}D^c \hat n_L \,.
\end{equation}
For all terms considered, $\mathcal J_L =0=\mathcal Q^-_{n,L}$ for all $L$ and $n \geqslant 2$ because the integrand is parity odd. For terms involving the sole quadrupole moment $M_{ij}$ at linear and quadratic orders and tails of the form $M \times M_{ij}$, the explicit results are 
\begin{subequations}\label{expressions}
\begin{align}
\mathcal{E} &= M -\frac{1}{5}G \int_{-\infty}^u M_{ij}^{(3)}M_{ij}^{(3)}, \\ 
\mathcal{P}_{ij} &=\frac{2}{5}G M_{ij}^{\text{rad}(2)}+\frac{2}{35}G^2\left[-2M_{ik}^{(2)}M_{jk}^{(3)}-5M_{ik}^{(1)}M_{jk}^{(4)}+M_{ik}M_{kj}^{(5)}\right]^{\mathrm{STF}}\,, \\
\mathcal{P}_{ijkl} &= \frac{1}{105}G^2\left[-34M_{ij}^{(2)}M_{kl}^{(3)}-21M_{ij}^{(1)}M_{kl}^{(4)}-7M_{ij}M_{kl}^{(5)}\right]^{\mathrm{STF}},\\
    \mathcal{K}_{ij} &= \frac{6}{5}G M_{ij}^{\text{rad}(1)}+\frac{6}{35}G^2 \left[ -M_{ik}^{(2)}M_{jk}^{(2)}-6M_{ik}^{(1)}M_{jk}^{(3)}+M_{ik}M_{jk}^{(4)}\right]^{\mathrm{STF}} \\
    \mathcal{K}_{ijkl} &= -\frac{2}{21}G^2\left[9M_{ij}^{(2)}M_{kl}^{(2)}+14M_{ij}^{(1)}M_{kl}^{(3)}+7M_{ij}M_{kl}^{(4)}\right]^{\mathrm{STF}}\,,\\
    \mathcal Q^+_{2,ij}&=  \frac{8}{5} G \biggl( M^{\text{rad}}_{ij}
 -   u M^{\text{rad}(1)}_{ij}
 + \frac{1}{2} u^2 M^{\text{rad}(2)}_{ij}\biggr)\nonumber\\
& + \frac{8}{35} G^2 \biggl[ -7 M^{(1)}_{ik} M^{(2)}_{jk} + M_{ik} M^{(3)}_{jk}
 +  u \biggl(7 M^{(2)}_{ik}M^{(2)}_{jk}
 + 6 M^{(1)}_{ik} M^{(3)}_{jk}
 -  M_{ik} M^{(4)}_{jk}\biggr)\nonumber\\
&\qquad\qquad-  \frac{1}{2} u^2 \biggl(2 M^{(2)}_{ik} M^{(3)}_{jk}
 + 5 M^{(1)}_{ik} M^{(4)}_{jk}
 -  M_{ik} M^{(5)}_{jk}\biggr)\biggl]^{\mathrm{STF}} \,,\\
    \mathcal Q^+_{3,ij}&=4G^2 MM_{ij}(-\mathcal T) \,,
\end{align}
\end{subequations}
where $M^{\text{rad}}_{ij}(u)$ contains the tail interactions of the form $M \times M_{ij}$: 
\begin{equation}
    M^{\text{rad}}_{ij}(u) = M_{ij}(u) +2 GM\int_{0}^{+\infty}\dd z \left[\ln\left(\frac{z}{2 b_0}\right) + \frac{11}{12}\right]M_{ij}^{(2)}(u-z) + \mathcal{O}(G^2)\,.
\end{equation}
These charges were already computed at $\mathcal{O}(G)$ in~\cite{Compere:2022zdz}. 
 In particular, the set of charges for $2\leqslant \ell \leqslant n-1$, obeying memory-less flux-balance laws, were shown to be vanishing at the linear order. Including the quadratic contributions, the first non-trivial such charge $\mathcal Q^+_{3,ij}$ gets a constant value, given by the product between the ADM mass and the quadrupole moment at the early time $\mathcal{-T}$ when the system is assumed to be stationary. In fact, $\mathcal Q^+_{3,ij}$ exactly matches with the Newman-Penrose charges as demonstrated in eqs.~\eqref{NP charge simplified} and~\eqref{NPcharge}. 

For the complementary set of charges with $\ell \geqslant n$, obeying the memory-full flux-balance laws, we have computed ${}_{(2)}\mathcal{E}^{ab}$ and ${}_{(3)}\mathcal{E}^{ab}$ in eqs.~\eqref{curlyE2} and~\eqref{curvyE3} explicitly. For quadrupole-quadrupole interactions, the non-vanishing charges can only have 0, 2 or 4 free indices, since they are traceless quantities only built from $M_{ij}M_{kl}$, with $u$ derivatives, $u$-dependent factors and contractions of indices. The remaining non-vanishing charges for quadrupole-quadrupole interactions are
\begin{subequations}\label{expressions2}
\begin{align}
  \mathcal Q^+_{2,ijkl}&=G^2 \biggl[ -4\Bigl(M^{(1)}_{ij}M^{(2)}_{kl}+M_{ij}M^{(3)}_{kl}\Bigr)+4 u \Bigl(M^{(2)}_{ij}M^{(2)}_{kl} +2 M^{(1)}_{ij}M^{(3)}_{kl}+M_{ij}M^{(4)}_{kl}\Bigr) \nonumber\\
  &\qquad\quad +u^2\biggl(-\frac{68}{7} M^{(2)}_{ij}M^{(3)}_{kl}-6 M^{(1)}_{ij}M^{(4)}_{kl}-2 M_{ij}M^{(5)}_{kl}\biggr) \biggl]^{\mathrm{STF}}\,, \\ 
   \mathcal Q^+_{3,ijkl}&=G^2 \biggl[ \frac{8}{3}M^{(1)}_{ij}M^{(1)}_{kl}-14 M_{ij}M^{(2)}_{kl}+u\biggl(\frac{26}{3}M^{(1)}_{ij}M^{(2)}_{kl}+14 M_{ij}M^{(3)}_{kl}\biggr)\nonumber\\ 
   &\qquad\quad+u^2 \biggl(-\frac{13}{3} M^{(2)}_{ij}M^{(2)}_{kl} -7(2 M^{(1)}_{ij}M^{(3)}_{kl}+M^{}_{ij}M^{(4)}_{kl})\biggl) \nonumber\\ 
   &\qquad\quad+u^3 \biggl( \frac{34}{3} M^{(2)}_{ij}M^{(3)}_{kl}+7 M^{(1)}_{ij}M^{(4)}_{kl}+\frac{7}{3}M^{}_{ij}M^{(5)}_{kl}\biggr)\biggl]^{\mathrm{STF}}\,.
\end{align}
\end{subequations}
The stationary limit is obtained straightforwardly.

\section{Conclusions}\label{Sec:concl}
The asymptotic Bondi-Sachs formalism is a convenient setup to study various gravita\-tional-wave observables, such as fluxes of conserved quantities or the canonical structure of radiative phase space in terms of the Bondi shear. In this setup, the Bondi shear, which characterizes the gravitational-wave strain, is not constrained by Einstein's equations and is thus free data. However, to obtain the Bondi shear for a given source, one needs to combine it with other wave generation methods. This paper, following~\cite{Blanchet:2020ngx}, continues the programme to incorporate results from the PN/MPM formalism into the Bondi-Sachs setup in a systematic way. This analysis allows us to infer several interesting properties of the spacetime geometry; in particular, 1) we highlight its global features, such as the peeling property or the late-time behaviour of the waveform, including memory and tail effects, and 2) we check the conservation of asymptotic BMS charges and their generalizations.  
%Interestingly, we clarify how the 10 Newman-Penrose charges are strictly conserved, despite the presence of radiation with arbitrary time dependence. In the following, we summarize our main results.
%

%
In section~\ref{sec:quadratic}, we retrace the computation of hereditary terms in the MPM formalism~\cite{Blanchet:1992br}. We also apply the algorithm introduced in~\cite{Blanchet:2020ngx} to transform into NU gauge the sector of the 2PM metric containing non-linear memory and secular losses. This first result is presented in eqs.~\eqref{eq:gmunu} whereas the asymptotic data, consisting of the Bondi mass and angular momentum aspects and the Bondi shear, are written in eqs.~\eqref{eq:mNaCab}. The second and main result of this work is the NU metric at 2PM order, limited to multipole interactions of the type monopole-quadrupole $M\times M_{ij}$ and quadrupole-quadrupole $M_{ij}\times M_{ij}$, including tail terms (already obtained in~\cite{Blanchet:2020ngx}) as well as non-linear memory and mass-loss terms. The NU metric for such interactions is given in eqs.~\eqref{eq:gmunuNUform}, where the Bondi mass and angular momentum aspects are reported in eqs.~\eqref{eq:Bondiaspects} [supplemented by the tail contributions, resp., of the first two eqs.~\eqref{Entail}], while the Bondi shear is written in eqs.~\eqref{HijTT}-\eqref{HijTTexp} [supplemented by the tail contribution of the third equation in eqs.~\eqref{Entail}]. Sub-leading contributions to the $uu$, $ui$ and $ij$ components of the NU metric are displayed, resp., in eqs.~\eqref{guusub}-\eqref{guisub}-\eqref{gabsub}, being the latter supplemented by tail contributions to the $ij$ components in eqs.~\eqref{Entail}.
We also remark the explicit proof of the re-summation of the infinite multipole series in eq.~\eqref{eq:closedform}, which has seldomly appeared in the previous literature, and some special treatment of hereditary terms explained in App.~\ref{app:A}.
In section~\ref{sec:IR}, based on the previous section, we discuss gravitational charges and their time evolution in the presence of radiation. One non-trivial test of our results is the explicit check (limited to the specific multipole interactions considered in this work) that Newman-Penrose charges are conserved; they turn out to be proportional to the ADM mass times the initial mass quadrupole moment. A third outcome is the (corrected) derivation of the dressed $n=3$ Bondi aspect in eq.~\eqref{curvyE3} and its flux in eq.~\eqref{F3dressed}. The flux-balance laws for $n=2,3$ were checked with the help of a computer code up to quadrupole-quadrupole interactions. Finally, we explicitly write the first $n\leqslant 3$ Bondi charges, \emph{i.e.}, supermomenta, super-Lorentz and $n=2,3$ celestial charges, in eqs.~\eqref{expressions} and eqs.~\eqref{expressions2} in terms of the mass quadrupole interactions.

\paragraph{Acknowledgments}
G.C. is Senior Research Associate of the F.R.S.-FNRS and acknowledges
support from the FNRS research credit J.0036.20F and the IISN convention 4.4503.15.
The work of R.O.~is supported by the R\'egion \^Ile-de-France within the DIM ACAV$^{+}$ project SYMONGRAV (Sym\'etries asymptotiques et ondes gravitationnelles). 
A.S. is supported by a Royal Society University Research Fellowship.
L.B.~and R.O.~ackowledge support from the Partenariat Hubert Curien within the Barrande mobility programme (project number $46771$VC).

\appendix
\section{Treatment of hereditary integrals}
\label{app:A}

\subsection{Radial integration at fixed retarded time}

In order to systematize the construction of the Newman-Unti metric, we develop general formulae for handling the hereditary tail integrals entering the $M_{ij}\times M_{ij}$ harmonic metric. At quadratic order, they all take the form\footnote{For simplicity in this Appendix we denote the coordinates by $(t,r)$, and $u\equiv t-r$, although they are really meant to be the harmonic coordinates we start with in our calculation.} 
\begin{align}\label{eq:Im}
	I_m(t,r) = \int_1^{+\infty} \dd x \, Q_{m}(x) F(t-r x)\,.
\end{align}
Here $Q_m(x)$, $m\in \mathbb{N}$, is the Legendre function of the second kind (with branch cut singularity from $-\infty$ to 1), and $F(u)$ is a quadratic product of time derivatives of quadrupole moments, which vanishes when $u\leqslant -\mathcal{T}$, so that the integration range is actually finite. Explicit expressions of the Legendre function are given below in~\eqref{eq:exprQm} and~\eqref{eq:Qmexplicit}. They show that the only possible convergence issue concerns the bound $x=1$, but the convergence is actually guaranteed by the local behaviour $Q_m(x) \sim -\frac{1}{2}\ln (x-1)$ when $x\to 1^+$.

The perturbation equations~\eqref{eq:quadordersim} we must solve in our approach are of the type $k^\mu \partial_\mu J_m = r^{-k} I_m$. Let us first treat the case $k=0$, and see later how we deal with other cases. We are thus looking for a function $J_m(t,r)$ such that
\begin{align}\label{eq:dJmdr}
	\frac{\partial J_m(u+r,r)}{\partial r} \biggl|_{u=\text{const}} = I_m(t,r) \,.
\end{align}
The problem we meet with the hereditary term~\eqref{eq:Im} is that the integral $J_m$ formally satisfying eq.~\eqref{eq:dJmdr}, holding $u=\text{const}$, reads $-\int_1^{+\infty} \dd x\, F^{(-1)}(t-r x)Q_m(x)/(x-1)$, plus a possible integration constant, irrelevant for the present discussion. Due to the factor $(x-1)^{-1}$, the integral diverges since $Q_m(x)/(x-1)$ is no longer locally integrable at the bound $x=1^+$. A way around is to isolate the non-integrable part of $Q_m(x)/(x-1)$ modulo a well-behaved function. In the previous paper~\cite{Blanchet:2020ngx}, we tackled the problem only on a ``case-by-case'' basis. Here, we present a more general method, based on the following formula for the Legendre function, which we shall prove at the end of this subsection:
\begin{subequations}\label{eq:formuleQm}
	\begin{align}
		&Q_m(x) - Q_0(x) + H_m = (x-1) \sum_{j=0}^{m-1} c_{mj} \,Q_j(x)\,,\label{eq:formuleQma}\\
		&\text{with}~~ c_{mj} \equiv (2j+1)\bigl(H_m - H_j\bigr)\,,
	\end{align}
\end{subequations}
where $H_m=\sum_{k=1}^{m}k^{-1}$ denotes the usual harmonic number and we recall that $Q_0(x)=1/2\ln[(x+1)/(x-1)]$. Plugging $Q_m(x)$ as deduced from~\eqref{eq:formuleQma} into $I_m$, we readily obtain
\begin{align}\label{eq:Imnew}
	I_m &= \sum_{j=0}^{m-1} c_{mj} \int_1^{+\infty} \!\dd x \, (x-1) Q_{j}(x) F(t-r x) + \frac{1}{r} \int_0^{+\infty} \!\dd \tau \left[\frac{1}{2} \ln\left(1+\frac{2r}{\tau}\right)- H_m\right]\!F(u-\tau) \,.
\end{align}
There is now an explicit factor $x-1$ in the integrand of the first term, so it can be directly integrated and the result will be a sum of hereditary integrals of the same structure as~\eqref{eq:Im}. As for the second term, where we have posed $x=1+\frac{\tau}{r}$, it yields upon integration a dilogarithm function $\text{Li}_2(z) \equiv - \int_0^z \dd s \ln(1-s)/s$. Hence we obtain
\begin{subequations}
	\begin{align}\label{eq:Jmnew}
		J_m &= - \sum_{j=0}^{m-1} c_{mj} \int_1^{+\infty} \!\dd x \, Q_{j}(x) F^{(-1)}(t-r x) \nn\\&\qquad\qquad 
		- \int_0^{+\infty} \!\dd \tau \left[\frac{1}{2}\text{Li}_2\left(-\frac{2r}{\tau}\right) + H_m\ln\left(\frac{r}{r_0}\right)\right]\!F(u-\tau) \,,
	\end{align}
plus a possible constant with respect to $r$. Here, $F^{(-1)}$ is the time-antiderivative of $F$, and $r_0$ is an arbitrary integration constant which can be chosen to be the Hadamard regularization scale of the MPM formalism. Alternatively,  by decomposing the logarithm  arising in eq.~\eqref{eq:Imnew} as $\ln [1+\tau/(2r)]-\ln[\tau/(2r)]$, coming back to the original variable $x$ in the integral of the first term and, finally, integrating with respect to $r$ for constant $u$ with the same procedure as for $\int_1^{+\infty} \!\dd x \, (x-1) Q_{j}(x) F(t-r x)$, we get
\begin{align}\label{eq:Imnewalt}
		J_m &= - \sum_{j=0}^{m-1} c_{mj} \int_1^{+\infty} \!\dd x \, Q_{j}(x) F^{(-1)}(t-r x) -\frac{1}{2}\int_0^{+\infty} \!\frac{\dd \tau}{\tau} \ln\left(1+\frac{\tau}{2r}\right)F^{(-1)}(u-\tau) \nn\\&\qquad\qquad\qquad\qquad + \frac{1}{4} \int_0^{+\infty} \!\dd \tau \, \ln^2 \left( \frac{\tau}{2r} \right) F(u-\tau) - H_m\ln\left(\frac{r}{r_0}\right)\!F^{(-1)}(u) \,.
\end{align}
\end{subequations}
 This expression makes explicit the appearance of $\ln r$ in the far-zone expansion, assuming that $F(u)$ and $F^{(-1)}(u)$ identically vanish in the remote past. To recover those logarithms from the form~\eqref{eq:Jmnew}, we refer to the behaviour of $\text{Li}_2(z)$ as $z\to - \infty$, which immediately follows from the relation $\text{Li}_2(z^{-1})= -\pi^2/6- \text{Li}_2(z) -1/2 \ln^2 (-z)$, valid for $z \not\in [0,+\infty[$. An integration by parts then shows that the results~\eqref{eq:Jmnew} and~\eqref{eq:Imnewalt} differ by a mere constant in $r$ as expected. In our calculation of the NU metric, following the algorithm~\eqref{eq:quadordersim}, all the logarithms of $r$ and associated constants, as well as all dilogarithm functions disappear in the end. 

We treat now the case of hereditary terms of the type $I_m/r^k$, more general than the one we just studied. Those terms could be handled by means of successive integrations by parts on $r$, but it is more convenient to provide a formula that allows decreasing the power $k$ in their pre-factors, until we get back to the case $k=0$.

The idea consists in integrating by parts the hereditary term~\eqref{eq:Im} over $x$. We introduce the anti-derivative of the Legendre function of the second kind that vanishes when $x \to 1^+$. Using the known identity $\frac{\dd}{\dd x}[Q_{m+1}(x) - Q_{m-1}(x)] = (2m+1)\, Q_m(x)$ for the Legendre function~\cite{Whittaker}, we get (for $m\geqslant 1$) 
\begin{align}
	\stackrel{(-1)}{Q}_{\!\!m}(x) \equiv \int_{1}^x \dd z \, Q_m(z) = \frac{1}{2m+1} \Bigl(Q_{m+1}(x) - Q_{m-1}(x) + H_{m+1} - H_{m-1} \Bigr)\,. 
\end{align}
Therefore, we find that an equivalent expression for $I_m$, which is obtained by integration by parts over $x$, reads 
\begin{align}\label{eq:power_reduction}
	I_m = \frac{r}{2m+1} \int_1^{+\infty} \dd x \,\Bigl[Q_{m+1}(x) - Q_{m-1}(x)\Bigr] F^{(1)}(t-r x) + \frac{H_{m+1} - H_{m-1}}{2m+1} \,F(u)\,.
\end{align}
The remaining hereditary integral is now endowed with an explicit extra factor $r$, so that, by iterating the process, we end up considering only hereditary integrals without prefactor $r^{-k}$, \emph{i.e.}, the simpler case studied before. 

In our calculation of the NU metric, we also need to differentiate $I_m$ with respect to $r$ at $u=$ const. For completeness, we give the formula for the radial derivative of $I_m$, which relies on the fact that the Legendre function satisfies the recurrence formula $(2m+1)\, x \, Q_m(x) = (m+1) Q_{m+1}(x)+ m Q_{m-1}(x)$ (for $m\geqslant 1$)~\cite{Whittaker}:
\begin{align}\label{eq:dImdr}
	\frac{\partial I_m}{\partial r} \biggl|_{u=\text{const}} \!\!= - \frac{1}{2m+1} \int_1^{+\infty} \!\dd x \, \Bigl[ (m+1) Q_{m+1}(x) - (2m+1) Q_{m}(x) + m Q_{m-1}(x) \Bigr] F^{(1)}(t-r x)\,.
\end{align}
Since the latter integral is a radial derivative, it can be directly integrated and, indeed, we see that the combination of Legendre functions in the integrand of~\eqref{eq:dImdr} is locally integrable when $x\to 1^+$.

%\subsection{Proof of the formula \texorpdfstring{\eqref{eq:formuleQm}}{}}
%\label{sec:proofformula}

\paragraph{Proof of the formula \texorpdfstring{\eqref{eq:formuleQm}}{}.} The coefficients $c_{mj}$ in this formula are those entering the decomposition of the polynomial of degree $m-1$
\begin{align}\label{eq:decompPm}
	\frac{P_m(x)-1}{x-1} = \sum_{j=0}^{m-1} c_{mj} \,P_j(x)\,,
\end{align}
into a sum of ordinary Legendre polynomials, where we recall that $P_m(1)=1$. The coefficients are readily computed as (for $j\leqslant m$)
\begin{align}\label{eq:computecoeff}
	c_{mj} &= \frac{2j+1}{2} \int_{-1}^{1} \dd x \,\frac{P_m(x)-1}{x-1} P_j(x) = \frac{2j+1}{2}	\lim_{z\to 1}\int_{-1}^{1} \dd x \,\frac{P_m(x)-1}{x-z} P_j(x) \nn\\ &= (2j+1) \lim_{z\to 1} \,\Bigl[ - P_j(z)Q_m(z) + Q_j(z)\Bigr] = (2j+1)\bigl(H_m - H_j\bigr)\,.
\end{align}
Notice the side results 
\begin{align}\label{eq:sideresults}
	\sum_{j=0}^{m-1} c_{mj} = \frac{m(m+1)}{2}\,,\qquad\qquad\sum_{j=0}^{m-1} c_{mj}H_j = \frac{m+1}{2}\bigl( m H_m-m +1\bigr)\,,
\end{align}
which may be derived by commuting the sum on $j$ with the one originating from the definition of harmonic numbers. Next, we know that the Legendre function reads
\begin{align}\label{eq:exprQm}
	Q_m(x) = \frac{1}{2} P_m(x)\ln\left(\frac{x+1}{x-1}\right) - W_{m-1}(x) \,,
\end{align}
where $W_{m-1}(x)$ is a polynomial of degree $m-1$ which consists of the positive and zero powers of $x$ in the expansion of $\frac{1}{2} P_m(x)\ln(\frac{x+1}{x-1})$ in descending powers of $x$, \emph{i.e.}, when $x\to+\infty$~\cite{Whittaker}. Notably, we have $W_{-1}(x)=0$ and $W_{m-1}(1)=H_m$. Using~\eqref{eq:exprQm} together with the decomposition~\eqref{eq:decompPm} yields
\begin{subequations}
	\begin{align}
		\frac{Q_m(x) - Q_0(x) + H_m}{x-1} &= \sum_{j=0}^{m-1} c_{mj} \,Q_j(x) + R_{m-2}(x)\,,\label{eq:formuleR}\\ \text{where}~~ R_{m-2}(x) &= \sum_{j=0}^{m-1} c_{mj} \,W_{j-1}(x) - \frac{W_{m-1}(x) - W_{m-1}(1)}{x-1}\,.
	\end{align}
\end{subequations}
The point is that $R_{m-2}(x)$ is a polynomial, with degree $m-2$. However, the existence of non-vanishing coefficients in $R_{m-2}(x)$ is incompatible with taking the limit when $x\to +\infty$ on both sides of eq.~\eqref{eq:formuleR}, since $Q_m(x)$ tends to zero like $1/x^{m+1}$ when $x\to+\infty$ (see \emph{e.g.}, eqs.~(A3) in~\cite{MQ4PN_jauge}). We conclude that this polynomial must be identically zero: $R_{m-2}(x)\equiv 0$, hence our formula~\eqref{eq:formuleQm} is proven. Moreover, we see that~\eqref{eq:formuleQm} is equivalent to
\begin{align}\label{eq:Wm}
	 \frac{W_{m-1}(x) - W_{m-1}(1)}{x-1} = \sum_{j=0}^{m-1} c_{mj} \,W_{j-1}(x)\,.
\end{align}

\subsection{Far-zone expansion}

In order to determine the far-zone behaviour of the hereditary integral $I_m$, we write it in terms of the variable $\tau=r (x-1)$ as
\begin{align} \label{eq:Im_tau}
	I_m = \frac{1}{r} \int_0^{+\infty} \dd \tau\, Q_m \left(
	1+\frac{\tau}{r} \right) F(u-\tau)\, .
\end{align}
Since the integration range is actually a bound interval, we can substitute to $Q_m(x)$ its asymptotic expansion near $x\to 1^+$ for large enough $r$. This expansion is derived from the following suitable representation of the Legendre function:
\begin{align}\label{eq:Qmexplicit}
	Q_m(x) = \sum_{j=0}^m \frac{(m+j)!}{(m-j)! j!^2}
	\left(\frac{x-1}{2}\right)^j \left[\frac{1}{2} \ln
	\left(\frac{x+1}{x-1}\right) + H_j - H_m 
	\right] \, .
\end{align}
Expanding $\ln(\frac{x+1}{2})$ when $x\to 1^+$, commuting the
sums, and resorting to standard resummation techniques including the
identification of hypergeometric functions,\footnote{In this instance, one may resort to the identity
	$${}_4 F_3(-n,1,1,x;2,y,x-y-n+1;1) = \frac{(y-1)(x-y-n)}{(n+1)(x-1)}\bigl[ -\psi(y-1)
	+ \psi(x-y+1) + \psi(y+n)- \psi(x-y-n)\bigr] \,,$$
	valid for $n\in \mathbb{N}$, where ${}_4 F_3$ is a generalized hypergeometric function and $\psi$ is the digamma function.} we get
\begin{align} \label{eq:Qm_exp}
	Q_m(x) &\overset{x\to 1^+}{\sim} \,\frac{1}{2} \sum_{j=0}^m
	\frac{(m+j)!}{(m-j)! j!^2} \left[2 H_j - 
	H_{m+j} - H_{m-j} - \ln \left( \frac{x-1}{2}\right) \right]
	\left(\frac{x-1}{2}\right)^j \nonumber \\ &\qquad
	+ \frac{1}{2} \left(\frac{x-1}{2}\right)^{m+1} \sum_{j=0}^{+\infty} (-)^j
	\,\frac{j! (2m+j+1)!}{(m+j+1)!^2} \left(\frac{x-1}{2}\right)^j\, .
\end{align}
An alternative form of this expansion is given by eq.~(4.6) in~\cite{TLB22}. Having inserted the expression \eqref{eq:Qm_exp} in eq.~\eqref{eq:Im_tau}, we permute the sums and the integrals, perform series of integration by parts to separate ``instantaneous'' terms (but which involve anti-derivatives of the function $F$) from logarithmic kernel hereditary integrals. We finally obtain the expansion, for $r\to +\infty$ at $u=$ const,
\begin{align}
	I_m &\overset{r\to +\infty}{\sim}  \,\sum_{j=0}^m
	\frac{(m+j)!}{(2r)^{j+1} (m-j)! j!} \bigg[\left(H_j - 
	H_{m+j} - H_{m-j}  \right)
	\!\stackrel{(-j-1)}{F}\!(u) -
	\int_0^{+\infty} \dd \tau \,  \ln
	\left( \frac{\tau}{2r} \right)  \!\stackrel{(-j)}{F}\!(u-\tau)\bigg]
	\nonumber \\ &\qquad +  \sum_{j=m+1}^{+\infty} (-)^{j+m+1}
	\frac{(j-m-1)! (j+m)!}{(2r)^{j+1}j!}
	\!\stackrel{(-j-1)}{F}(u)\, . 
\end{align}
For the expansion of $I_m$ in the near zone, \emph{i.e.}, $r\to 0$ with $u$ or $t$ fixed, we refer the reader to App.~A of~\cite{MQ4PN_jauge} .

The appearance of arbitrarily high order anti-derivatives reflects the non-locality of $I_m$. Note also the presence of terms proportional to $\ln r/r^{j+1}$ with $j=0, \cdots, m$. Those logarithms, however, have to disappear from the NU metric components by construction. Moreover, for the quadrupole-quadrupole interaction $M_{ij}\times M_{ij}$, it turns out that all hereditary integrals $I_{m}$ cancel each other once they have been transformed with the help of the power reduction formula~\eqref{eq:power_reduction} so as to bear the same common pre-factor $r^{-k_0}$ for some chosen $k_0$. The $M_{ij}\times M_{ij}$ non-local terms of the NU metric are thus of pure memory type and their far-zone expansion involves only a finite number of terms, which contrasts with the situation in harmonic coordinates.

\section{Dressed Bondi aspects for \texorpdfstring{\boldmath$M_{ij}\times M_{ij}$ \unboldmath}{} interactions}
\label{app:B}

Including only mass monopole and quadrupoles we have 
\begin{subequations}
\begin{align}
    C_{bc} &= 2G e^{i}_{\langle b}e^{j}_{c\rangle} {M}_{ij}^{(2)}+\mathcal{O}(G^2)\,,\\
    D^c C_{bc} &= - 4G e^{i}_b n^j {M}_{ij}^{(2)}+\mathcal{O}(G^2)\,, \\
    D_a D^c C_{bc} &= 4G \left(n_i n_j {M}_{ij}^{(2)}\gamma_{ab} - e^{i}_a e^{j}_b {M}_{ij}^{(2)} \right) + \mathcal{O}(G^2)\,. \label{DDC}
\end{align}
\end{subequations}
Notice that eq.~\eqref{DDC} is symmetric in $(ab)$, hence the identity $D_{[a}D^c C_{b]c} = \mathcal{O}(G^2)$. 
We then deduce from this\footnote{It is useful to observe that, from $C_{ab}=e^i_{\langle a}e^j_{b \rangle}H_{ij}^{TT}$ [see~\eqref{HijTT}], one has
\begin{align*}
    C_{ab} &= 
    e^{k}_{\langle a} e^{l}_{b\rangle} \left(2 U_{kl} - n_m n_p \eps_{mn(k}V_{l)np} + \frac{1}{6}n_p n_q U_{klpq}\right)\,, \\
      D^c C_{bc} &= -2n^{k}e^{l}_b \left(2U_{kl}+\frac{1}{4}n_p n_q U_{klpq} -\frac{5}{4}n_m \eps_{mnl}V_{knp}n_p\right)\,,\\
      D_{[a} D^c C_{b]c} &= 5 \left(e^k_{[a} e^l_{b]} n^m  + \frac{1}{2}e^m_a e^l_b n^k\right)\eps_{mnl} V_{knp}n_p\,,\label{Celec} \\
      D^bD_{[a} D^c C_{b]c} &= 15 \ e^l_a n^{m}\eps_{mnl}V_{knp}n_{k}n_p\,.
\end{align*}
}
\begin{subequations}
\begin{align}
  D_a \left(C^2 \right) &= -8 G^2 e^i_a {M}_{ij}^{(2)} \left(2  {M}_{jk}^{(2)}-n_jn_l {M}_{lk}^{(2)}\right)n_k+\mathcal{O}(G^3)\,, \\ 
    C_a^{\;b}D^cC_{bc} &= -4G^2 e_a^i {M}_{ij}^{(2)} \left(2 {M}_{jk}^{(2)} - 
     n_j n_l {M}_{lk}^{(2)} \right)n_k +\mathcal{O}(G^3)\,,\\
     D^bD_{[a} D^c C_{b]c} &= 15 G^2\ e^l_a n^{m} \eps_{lmj}n^{i}\eps_{pq\langle i}\left(\frac{1}{2}  M^{(1)}_{j\underline{p}} M^{(4)}_{k\rangle q}
		-  \frac{1}{10} M_{j\underline{p}} M^{(5)}_{k\rangle q}\right)n_k + \mathcal{O}(G^3)\,,
\end{align}
\end{subequations}
while $D_a m$ is immediately computed from eq.~\eqref{mexpl}. 

Resorting to a computer algebra software, we now include all quadratic-quadratic interactions to obtain
\begin{align}
\mathcal{N}_{i} &=G \biggl [3 M_{i}
 - 3 u P_{i}
 + \hat{n}^{m} \left(6 M^{(1)}_{im}
 - 6 u M^{(2)}_{im}\right)\biggl]^{\text{T}}
 + G^2 \biggl\{\hat{n}^{m} \biggl [\frac{12}{5} \int_{-\infty}^{u} \!\dd v M^{(2)}_{n[m}(v) M^{(3)}_{i]n}(v)\nonumber\\
& -  \frac{264}{35} M^{(1)}_{n(m} M^{(3)}_{i)n}
 -  \frac{12}{35} M_{n(m} M^{(4)}_{i)n} + u \biggl(\frac{264}{35} M^{(2)}_{n(m} M^{(3)}_{i)n}
 + \frac{276}{35} M^{(1)}_{n(m} M^{(4)}_{i)n}
 + \frac{12}{35} M_{n(m} M^{(5)}_{i)n}\biggr)\biggr]\nonumber\\
& + \hat{n}^{mnj} \biggl[-15 M^{(2)}_{im} M^{(2)}_{nj}
 - 15 M^{(1)}_{mn} M^{(3)}_{ij}
 - 6 M^{(1)}_{im} M^{(3)}_{nj}
 -  \frac{9}{2} M_{mn} M^{(4)}_{ij}
 - 6 M_{im} M^{(4)}_{nj}\nonumber\\
& + u \biggl(51 M^{(2)}_{m(n} M^{(3)}_{i)j}
 + \frac{39}{2} M^{(1)}_{mn} M^{(4)}_{ij}
 + 12 M^{(1)}_{im} M^{(4)}_{nj}+ \frac{9}{2} M_{mn} M^{(5)}_{ij}
 + 6 M_{im} M^{(5)}_{nj}\biggr)\biggr]\biggr\}^{\text{T}},\\
 %%%%%%%%%%%%%%%%%%%%%%%%%%%%%%%%%%%
\stackrel[(2)]{}{\mathcal{E}_{ij}} &= G \Bigl(2 M_{ij}
 - 2 u M^{(1)}_{ij}
 + u^2 M^{(2)}_{ij}\Bigr)^{\text{TT}}
 + G^2 \biggl\{- \frac{8}{3} M^{(1)}_{im} M^{(2)}_{jm}
 -  \frac{52}{63} M_{im} M^{(3)}_{jm}
 + u \biggl(2 M^{(2)}_{im} M^{(2)}_{jm}\nonumber\\
& + \frac{92}{21} M^{(1)}_{im} M^{(3)}_{jm}
 + \frac{22}{21} M_{im} M^{(4)}_{jm}\biggr)
 + u^2 \biggl(- \frac{74}{21} M^{(2)}_{im} M^{(3)}_{jm}
 -  \frac{19}{7} M^{(1)}_{im} M^{(4)}_{jm}
 -  \frac{11}{21} M_{im} M^{(5)}_{jm}\biggr)\nonumber\\
& + \hat{n}^{mn} \biggl [-3 M^{(1)}_{mn} M^{(2)}_{ij}
 - 8 M^{(1)}_{im} M^{(2)}_{jn}
 + \frac{1}{2} M^{(1)}_{ij} M^{(2)}_{mn}
 -  \frac{4}{3} M_{mn} M^{(3)}_{ij}
 -  \frac{22}{3} M_{im} M^{(3)}_{jn}\nonumber\\
& -  \frac{11}{6} M_{ij} M^{(3)}_{mn}
 + u \bigl(9 M^{(2)}_{im} M^{(2)}_{jn}
 + \frac{3}{2} M^{(2)}_{ij} M^{(2)}_{mn}
 + 5 M^{(1)}_{mn} M^{(3)}_{ij}
 + 14 M^{(1)}_{im} M^{(3)}_{jn}
 + 2 M^{(1)}_{ij} M^{(3)}_{mn}\nonumber\\
& + \frac{3}{2} M_{mn} M^{(4)}_{ij}
 + 7 M_{im} M^{(4)}_{jn}
 + 2 M_{ij} M^{(4)}_{mn}\bigr)
 + u^2 \biggl(- \frac{17}{4} M^{(2)}_{mn} M^{(3)}_{ij}
 - 17 M^{(2)}_{im} M^{(3)}_{jn}\nonumber\\
& -  \frac{17}{4} M^{(2)}_{ij} M^{(3)}_{mn}
 -  \frac{13}{4} M^{(1)}_{mn} M^{(4)}_{ij}
 -  \frac{21}{2} M^{(1)}_{im} M^{(4)}_{jn}
 - 2 M^{(1)}_{ij} M^{(4)}_{mn}
 -  \frac{3}{4} M_{mn} M^{(5)}_{ij}
 -  \frac{7}{2} M_{im} M^{(5)}_{jn}\nonumber\\
& -  M_{ij} M^{(5)}_{mn}\biggr)\biggl]\biggl\}^{\text{TT}},\\
%%%%%%%%%%%%%
%%%%%%%%%%%%%
%%%%%%%%%%%%%
%%%%%%%%%%%%%
\stackrel[(3)]{}{\mathcal{E}_{ij}}&= G^2 \biggl\{-2 M_{im} M^{(2)}_{jm}
 + u \hat{n}^{m} \bigl(-2 M_{m} M^{(2)}_{ij}
 - 4 M_{i} M^{(2)}_{jm}\bigr)
 + u \biggl(\frac{13}{3} M^{(1)}_{im} M^{(2)}_{jm}
 + \frac{41}{9} M_{im} M^{(3)}_{jm}\biggr)\nonumber\\
& + u^2 \biggl(-2 M^{(2)}_{im} M^{(2)}_{jm}
 -  \frac{14}{3} M^{(1)}_{im} M^{(3)}_{jm}
 -  \frac{7}{3} M_{im} M^{(4)}_{jm}\biggr)
 + u^3 \biggl(\frac{34}{9} M^{(2)}_{im} M^{(3)}_{jm}
 + \frac{7}{3} M^{(1)}_{im} M^{(4)}_{jm}\nonumber\\
& + \frac{7}{9} M_{im} M^{(5)}_{jm}\biggr)
 + \hat{n}^{mn} \biggl [6 M^{(1)}_{im} M^{(1)}_{jn}
 + M^{(1)}_{ij} M^{(1)}_{mn}
 -  \frac{15}{4} M_{mn} M^{(2)}_{ij}
 -  \frac{57}{2} M_{im} M^{(2)}_{jn}
 -  \frac{9}{2} M_{ij} M^{(2)}_{mn}\nonumber\\
& + u \biggl(\frac{9}{2} M^{(1)}_{mn} M^{(2)}_{ij}
 + 13 M^{(1)}_{im} M^{(2)}_{jn}
 + \frac{21}{4} M^{(1)}_{ij} M^{(2)}_{mn}
 + \frac{17}{3} M_{mn} M^{(3)}_{ij}
 + \frac{74}{3} M_{im} M^{(3)}_{jn}\nonumber\\
& + \frac{77}{12} M_{ij} M^{(3)}_{mn}\biggr)
 + u^2 \biggl(- \frac{27}{4} M^{(2)}_{im} M^{(2)}_{jn}
 -  \frac{37}{8} M^{(2)}_{ij} M^{(2)}_{mn}
 -  \frac{29}{4} M^{(1)}_{mn} M^{(3)}_{ij}
 -  \frac{49}{2} M^{(1)}_{im} M^{(3)}_{jn}\nonumber\\
& - 5 M^{(1)}_{ij} M^{(3)}_{mn}
 -  \frac{23}{8} M_{mn} M^{(4)}_{ij}
 -  \frac{49}{4} M_{im} M^{(4)}_{jn}
 -  \frac{13}{4} M_{ij} M^{(4)}_{mn}\biggr)
 + u^3 \biggl(\frac{119}{24} M^{(2)}_{mn} M^{(3)}_{ij}\nonumber\\
& + \frac{119}{6} M^{(2)}_{im} M^{(3)}_{jn}
 + \frac{119}{24} M^{(2)}_{ij} M^{(3)}_{mn}
 + \frac{27}{8} M^{(1)}_{mn} M^{(4)}_{ij}
 + \frac{49}{4} M^{(1)}_{im} M^{(4)}_{jn}
 + \frac{11}{4} M^{(1)}_{ij} M^{(4)}_{mn}\nonumber\\
& + \frac{23}{24} M_{mn} M^{(5)}_{ij}
 + \frac{49}{12} M_{im} M^{(5)}_{jn}
 + \frac{13}{12} M_{ij} M^{(5)}_{mn}\biggr)\biggl]\biggl\}^{\text{TT}}.
\end{align}

\bibliographystyle{JHEP}
\bibliography{ref_multipoles}
%--------------------------------------------------------------------------------------------------
%--------------------------------------------------------------------------------------------------
\end{document}